\documentclass{jpp}

\newcommand{\apj}{\mbox{\it Astrophysical Journal}}

\newcommand{\ssr}{\mbox{\it Space Science Rev.}}
\newcommand{\pre}{\mbox{\it Phys. Rev. E}}

\usepackage{color}
\usepackage{graphicx}
\usepackage{natbib}

\ifCUPmtlplainloaded \else
  \checkfont{eurm10}
  \iffontfound
    \IfFileExists{upmath.sty}
      {\typeout{^^JFound AMS Euler Roman fonts on the system,
                   using the 'upmath' package.^^J}%
       \usepackage{upmath}}
      {\typeout{^^JFound AMS Euler Roman fonts on the system, but you
                   dont seem to have the}%
       \typeout{'upmath' package installed. JPP.cls can take advantage
                 of these fonts, if you use 'upmath' package.^^J}%
      }
  \else
  \fi
\fi

\ifCUPmtlplainloaded \else
  \checkfont{msam10}
  \iffontfound
    \IfFileExists{amssymb.sty}
      {\typeout{^^JFound AMS Symbol fonts on the system, using the
                'amssymb' package.^^J}%
       \usepackage{amssymb}%

      }{}
  \fi
\fi

\ifCUPmtlplainloaded \else
  \IfFileExists{amsbsy.sty}
    {\typeout{^^JFound the 'amsbsy' package on the system, using it.^^J}%
     \usepackage{amsbsy}}
    {}
\fi

\newsavebox{\astrutbox}
\sbox{\astrutbox}{\rule[-5pt]{0pt}{20pt}}

\title[Density jump for oblique collisionless shocks in pair plasmas: allowed solutions]{Density jump for oblique collisionless shocks in pair plasmas: allowed solutions}

\author[A. Bret and R. Narayan]%
{Antoine Bret$^{1,2}$, Ramesh Narayan$^{3,4}$%
  \thanks{Email address for correspondence: antoineclaude.bret@uclm.es}
}

\affiliation{$^1$ETSI Industriales, Universidad de Castilla-La Mancha, 13071 Ciudad Real, Spain\\[\affilskip]
$^2$Instituto de Investigaciones Energ\'{e}ticas y Aplicaciones Industriales, Campus Universitario de Ciudad Real, 13071 Ciudad Real, Spain\\[\affilskip]
$^3$ Center for Astrophysics | Harvard and Smithsonian, Harvard University, 60 Garden St., Cambridge, MA 02138, USA\\
$^4$Black Hole Initiative at Harvard University, 20 Garden St., Cambridge, MA 02138, USA
}

\date{?; revised ?; accepted ?. - To be entered by editorial office}

\begin{document}

\maketitle

\begin{abstract}
Shockwaves in plasma are usually dealt with using Magnetohydrodynamics (MHD). Yet, MHD entails the  assumption of a short mean free path, which is not fulfilled in a collisionless plasma. Recently, for pair plasmas, we devised a model allowing to account for kinetic effects within an MHD-like formalism. Its relies on an estimate of the anisotropy generated when crossing the front, with a subsequent assessment of the stability of this anisotropy in the downstream.  We solved our model for parallel, perpendicular and switch-on shocks. Here we bridge between all these cases by treating the problem of an arbitrarily, but coplanar, oriented magnetic field. Even though the formalism presented is valid for anisotropic upstream temperatures, only the case of a cold upstream is solved. We find extra solutions which are not part of the MHD catalog, and a density jump that is notably less in the quasi parallel, highly magnetized, regime. Given the complexity of the calculations, this work is mainly devoted to the presentation of the mathematical aspect of our model. A forthcoming article will be devoted to the physics of the shocks here defined.
\end{abstract}

\maketitle

\section{Introduction}
Shock waves in plasmas are typically analysed using the tools of Magnetohydrodynamics (MHD). Hence, the jump conditions derived rely on two assumptions: 1) that collisions are frequent enough to establish an isotropic pressure, both upstream and downstream, and 2) that all the matter upstream passes to the downstream, together with the momentum and the energy it carries (\cite{gurnett2005} \S 5.4.4, \cite{Goedbloed2010} chapters 2 \& 3, or  \cite{TB2017} \S 13.2).

It turns out that in collisionless plasmas, where the mean free path is much larger than the size of the system, shock front included, these two assumptions may not be fulfilled. Regarding the second one, it has been known for long that collisionless shocks can accelerate particles which escape the ``Rankine-Hugoniot (RH) budget'' and modify the jump conditions \citep{Berezhko1999}. As for the first assumption, namely that pressures are isotropic, it is still valid in a collisionless un-magnetized plasmas since in such plasmas, the Weibel instability ensures isotropic pressures are unstable \citep{Weibel,SilvaPRE2021}.

Yet, still in a collisionless plasma, an external magnetic field can stabilize an anisotropy, invalidating the second assumption \citep{Hasegawa1975,Gary1993}. This has been clearly proved by \emph{in situ} measurement in the solar wind \citep{BalePRL2009,MarucaPRL2011,SchlickeiserPRL2011}. The present work is about departures from MHD predictions stemming from the violation of the second assumption. Departures stemming from the violation of the first one, namely accelerated particles escaping the RH budget, will not be addressed here (see \cite{BretApJ2020} for a review).

Assuming an isotropic upstream, how could any anisotropy develop downstream? Simply through an anisotropy that would be triggered at the front crossing, and then maintained stable in the downstream by means of an external magnetic field. Such is the scenario we have been contemplating in a series of recent articles on parallel, perpendicular and switch-on shocks \citep{BretJPP2018,BretPoP2019,BretLPB2020,BretJPP2022}.

In our model, the plasma is compressed anisotropically  when it crosses the front. Then, depending on the resulting anisotropy degree, the field can sustain the anisotropy in the downstream, or not. Note that for the parallel case, our model has been successfully tested against Particle-In-Cell (PIC) simulations in \cite{Haggerty2022}.

The present work aims at bridging between all the previously treated cases. We shall therefore consider the general case of an oblique shock, where the upstream magnetic field makes an arbitrary angle with the shock normal.

\begin{figure}
\begin{center}
 \includegraphics[width=\textwidth]{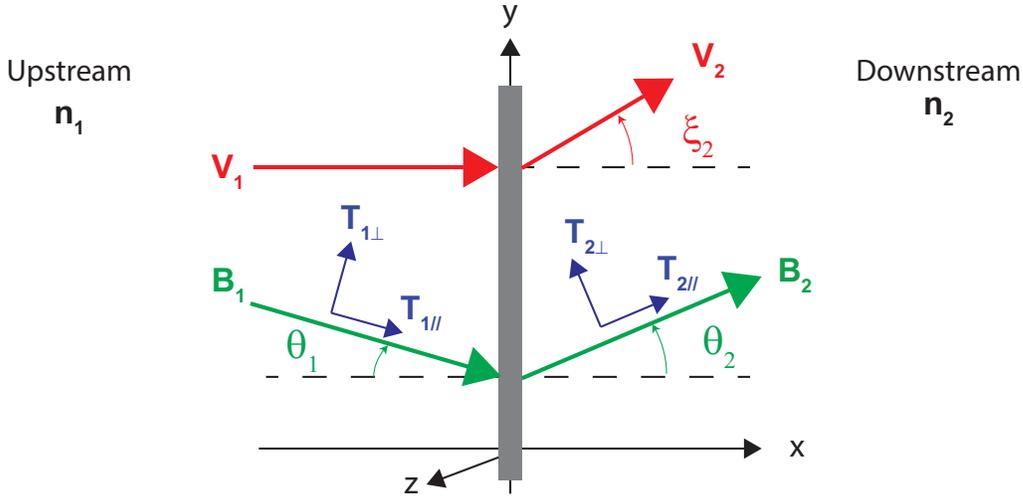}
\end{center}
\caption{System considered. The upstream magnetic field $\mathbf{B}_1$ makes an angle $\theta_1$ with the shock normal. The downstream field $\mathbf{B}_2$ and velocity $\mathbf{v}_2$ make angles $\theta_2$ and $\xi_2$ respectively with the shock normal. We work in the reference frame where the shock is stationary and the upstream velocity $\mathbf{v}_1$ is normal to the front ($\xi_1=0$). The upstream has density $n_1$ and  temperatures $T_{1\parallel}, T_{1\perp}$, parallel and perpendicular to the upstream field $\mathbf{B}_1$. The downstream has density $n_2$ and  temperatures $T_{2\parallel}, T_{2\perp}$, parallel and perpendicular to the downstream field $\mathbf{B}_2$.  The parallel and perpendicular directions are therefore defined with respect to the local magnetic field.\\
Even though the equations presented in section \ref{sec:equations} can be applied to an anisotropic upstream, the model is only solved for $T_{1\parallel}= T_{1\perp}=0$.}\label{fig:system}
\end{figure}

The system considered is pictured on figure \ref{fig:system}. Sub-indices ``1'' and ``2'' refer to the upstream and the downstream respectively.  We work in the reference frame where the upstream velocity $\mathbf{v}_1$ is normal to the front. The upstream magnetic field $\mathbf{B}_1$ makes an arbitrary angle $\theta_1 \neq 0$ with the shock normal, contrary to \cite{BretJPP2018,BretJPP2022} where $\theta_1=0$, and to \cite{BretPoP2019} where $\theta_1=\pi/2$. The fields $\mathbf{B}_{1,2}$ and the velocities $\mathbf{v}_{1,2}$ are assumed coplanar.

Even though the formalism presented is valid for anisotropic downstream \emph{and} upstream temperatures, we shall restrict to $T_{1\parallel}= T_{1\perp}=0$ when solving it.

Also, we consider a plasma of electron/positron pairs. The identity of the mass of both species allows us to deal with only one parallel and one perpendicular temperature in the downstream, as it has been found that in collisionless shocks, species of different mass are heated differently \citep{Feldman1982,Guo2017,Guo2018}.

As the reader will realize, even for a coplanar geometry with $T_{1\parallel}= T_{1\perp}=0$, the forthcoming algebra is quite involved. For this reason, the present work is mainly devoted to the algebraic resolution of our model for the oblique case. We write down the conservation equations and explain how to solve them symbolically. We also explain how these solutions fit with each others within the rules of our model. Yet, as known even for MHD, listing the solutions of the equations does not provide the full picture of the shock physics, as some solutions which do satisfy the MHD conservation equations could eventually be nonphysical \citep{Kennel1990,Komissarov1997,Wu2003,Kulsrud2005,Goedbloed2008,Delmont2011}. An assessment of the physical relevance of our solutions will be presented in a forthcoming article. Here, we shall focus on the mathematical solutions of our model.

This article is structured as follow. In section \ref{sec:method}, we explain our model, emphasizing how we bridge between our previous treatments of the parallel and the perpendicular cases.  In particular, we introduce ``Stage 1'' and ``Stage 2'' which are supposed to be 2 stages of the kinetic history of the plasma. In section \ref{sec:equations}, we introduce the conservation equations for anisotropic temperatures, together with the dimensionless variables used in the sequel. In sections \ref{sec:stage1}, \ref{sec:S2fire} and \ref{sec:S2mirror}, Stages 1 and 2  are studied separately. Then in section \ref{sec:together}, we explain how they relate to each other in order to fully characterize the shock within our model for any field obliquity $\theta_1$.

\section{Method}\label{sec:method}
Although the method used to deal with the oblique case has been explained in \cite{BretJPP2022}, we here outline it for completeness.

Consider an upstream with temperature $T_{1\parallel}$ and $T_{1\perp}$. If the crossing of the front could be fully described by the isentropic Vlasov equation (\cite{LandauKinetic}, \S 27), the  downstream temperatures could be related to the other quantities through the expressions derived in \cite{CGL1956},
\begin{eqnarray}\label{eq:CGL}
  T_{2\parallel} &=& T_{1\parallel} \left( \frac{n_2 B_1}{n_1 B_2}  \right)^2, \\
  T_{2\perp} &=& T_{1\perp} \frac{B_1}{B_2}. \nonumber
\end{eqnarray}
But the crossing of the front is not isentropic since in a shock, there is an entropy increase from the upstream to the downstream. As a consequence, temperatures increase by more than the amount specified by Eqs. (\ref{eq:CGL}), as found in the PIC simulations by \cite{Haggerty2022}. In both the parallel case ($\theta_{1,2}=0$) and the perpendicular case ($\theta_{1,2}=\pi/2$), we considered this excess goes into the temperature parallel to the motion, since the compression at the front can be considered to operate along this direction. As a consequence, the temperature parallel to the motion increases, while the temperature perpendicular to the motion remains constant.

Hence, denoting $T_\mathrm{entropy}$ the temperature correction due to entropy generation, we took for the parallel case,
\begin{eqnarray}\label{eq:CGLpara}
  T_{2\parallel} &=& T_{1\parallel} \left( \frac{n_2 B_1}{n_1 B_2}  \right)^2 + ~T_\mathrm{entropy}, \\
  T_{2\perp} &=& T_{1\perp} \frac{B_1}{B_2}, \nonumber
\end{eqnarray}
and for the perpendicular case,
\begin{eqnarray}\label{eq:CGLperp}
  T_{2\parallel} &=& T_{1\parallel}  \left( \frac{n_2 B_1}{n_1 B_2}  \right)^2, \\
  T_{2\perp} &=& T_{1\perp} \frac{B_1}{B_2}  + ~T_\mathrm{entropy}. \nonumber
\end{eqnarray}
In order to bridge between these two extremes, we now make the following \emph{ansatz}:
\begin{eqnarray}
  T_{2\parallel} &=& T_{1\parallel}  \left( \frac{n_2 B_1}{n_1 B_2}  \right)^2 + T_e\cos^2\theta_2, \label{eq:CGLoblique1}\\
  T_{2\perp} &=& T_{1\perp} \frac{B_1}{B_2} + \frac{1}{2}T_e\sin^2\theta_2, \label{eq:CGLoblique2}
\end{eqnarray}
where $T_e$ (subscript $e$ for $e$ntropy) will be determined by the conservation equations.

Physically, Eqs. (\ref{eq:CGLoblique1},\ref{eq:CGLoblique2}) are motivated by our hypothesis that the excess energy goes into a direction parallel to the upstream velocity, in analogy with our previous treatments of the parallel and perpendicular shocks sub-cases. Geometry is then used to divide the energy excess between $T_{2\parallel}$ and $T_{2\perp}$.

The scheme chosen in (\ref{eq:CGLoblique1},\ref{eq:CGLoblique2}) is the simplest one fulfilling the following conditions,
\begin{itemize}
  \item It correctly reduces to Eqs. (\ref{eq:CGLpara},\ref{eq:CGLperp}) for $\theta_2=0$ and $\theta_2=\pi/2$.
  \item All temperature excesses sum up to $T_e$.
  \item It guaranties the 2 downstream temperatures normal to the field $\mathbf{B}_2$ are equal, which is required by the Vlasov equation (\cite{LandauKinetic}, \S53).
\end{itemize}
Its relevance will have to be checked via PIC simulation, like \cite{BretJPP2018} has been checked in \cite{Haggerty2022}.

\bigskip

\begin{table}
\begin{center}
\begin{tabular}{l|c|r|l}
  Upstream  & Downstream in Stage 1                                  &      Stable?  ~~~~~~~~~   & End state of the downstream \\
  \hline
            & $T_{2\perp}$ and $T_{2\parallel}$       & Stable   $\rightarrow$         & Stage 1 \\
  $T_{1\parallel}= T_{1\perp}=0$ &  given by                                & Firehose unstable $\rightarrow$ & Stage-2-firehose \\
            & Eqs. (\ref{eq:CGLoblique1},\ref{eq:CGLoblique2}) & Mirror unstable  $\rightarrow$  & Stage-2-mirror
\end{tabular}
\end{center}
 \caption{Summary of the assumed kinetic history of the plasma as it crosses the front. Although the formalism presented in section \ref{sec:equations} allows for an anisotropic upstream, the model is only solved for $T_{1\parallel}=T_{1\perp}=0$.}\label{tab:cases}
\end{table}

The downstream temperatures after the front crossing are therefore given by Eqs. (\ref{eq:CGLoblique1},\ref{eq:CGLoblique2}). We refer to this state of the downstream as ``Stage 1''.
Depending of the strength of the downstream field $\mathbf{B}_2$, Stage 1 can be stable or unstable.

Previous analysis showed that Stage 1 can be firehose or mirror unstable. In case Stage 1 is firehose unstable, it migrates to the ``Stage-2-firehose'' state, on the firehose instability threshold where  \citep{Hasegawa1975,Gary1993,Gary2009},
\begin{equation}\label{eq:firehose}
A_2 \equiv \frac{T_{2\perp}}{T_{2\parallel}} = 1 - \frac{1}{\beta_{2\parallel}},
\end{equation}
with,
\begin{equation}\label{eq:beta2}
\beta_{2\parallel} = \frac{n_2 k_B T_{2\parallel}}{B_2^2/4\pi},
\end{equation}
where $k_B$ is the Boltzmann constant. In case Stage 1 is mirror unstable, it migrates to the ``Stage-2-mirror'' state, on  the mirror instability threshold where,
\begin{equation}\label{eq:mirror}
A_2  = 1 + \frac{1}{\beta_{2\parallel}}.
\end{equation}
At any rate, imposing condition (\ref{eq:firehose}) or (\ref{eq:mirror}) in the forthcoming conservation equations determines the state of the downstream. Our algorithm is summarized in Table \ref{tab:cases}.

\begin{figure}
\begin{center}
 \includegraphics[width=0.7\textwidth]{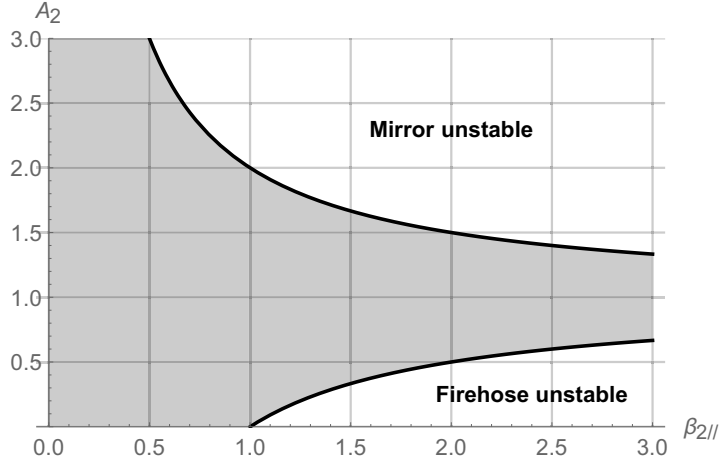}
\end{center}
\caption{Instability thresholds (\ref{eq:firehose},\ref{eq:mirror}) for the firehose and mirror instabilities. The system is stable in the shaded region and the instability domains do not overlap.}\label{fig:stab}
\end{figure}

The firehose instability reaches its maximum growth rate for $\mathbf{k}$ parallel to the field, while the mirror instability reaches its maximum growth rate for a $\mathbf{k}$ making an oblique angle with the field (\cite{Gary1993}, \S 7.2). The instability thresholds (\ref{eq:firehose},\ref{eq:mirror}) for the firehose and mirror instabilities are plotted on figure \ref{fig:stab}. Noteworthily, the instability domains do not overlap in the $(\beta_{2\parallel},A_2)$ plane so that the two instabilities cannot compete with each other.

\section{Conservation equations}\label{sec:equations}
The conservations equations for anisotropic temperatures were derived in \cite{Hudson1970,Erkaev2000,Genot2009}. They have been re-derived in \cite{BretJPP2022} with the present notations. They are formally valid even for anisotropic upstream temperatures, with $T_{1\parallel} \neq T_{1\perp}$. Writing them for $T_{1\parallel}=T_{1\perp}\equiv T_1$, they read,
\begin{eqnarray}
n_2 v_2 \cos  \xi _2  &=& n_1 v_1,  \label{eq:conser1}  \\
B_2 \cos \theta_2 &=& B_1 \cos \theta_1  ,  \label{eq:conser2} \\
B_2 v_2 \sin \theta_2 \cos \xi_2 - B_2 v_2 \cos \theta_2 \sin \xi_2 &=&  B_1 v_1 \sin \theta_1 , \label{eq:conser3} \\
\frac{B_2^2 \sin ^2\theta_2}{8 \pi }+n_2 k_B ( T_{\parallel 2} \cos ^2\theta_2 + T_{\perp 2} \sin ^2\theta_2) + m n_2 v_2^2 \cos ^2\xi_2
                                                  &=& \frac{B_1^2 \sin ^2\theta_1}{8 \pi }+n_2 k_B T_1 + m n_1 v_1^2 , \label{eq:conser4}\\
\mathcal{A} + m n_2 v_2^2 \sin \xi_2 \cos \xi_2 &=& - \frac{B_1^2 \sin \theta_1 \cos \theta_1}{4 \pi } , \label{eq:conser5}\\
\mathcal{A} v_2 \sin \xi_2 + \mathcal{B} + \mathcal{C} &=& m n_1 v_1 \left(\frac{5 k_B T_1 }{2 m}+\frac{B_1^2 \sin ^2\theta_1}{4 \pi  m n_1}+\frac{v_1^2}{2}\right), \label{eq:conser6}
\end{eqnarray}
where
\begin{eqnarray}
\mathcal{A} &=& \sin \theta_2 \cos \theta_2 n_2 k_B \left(T_{\parallel 2}-T_{\perp 2}\right) - \frac{B_2^2}{4 \pi }\sin \theta_2 \cos \theta_2, \nonumber\\
\mathcal{B} &=& v_2 \cos ^2\theta_2 \cos \xi_2 n_2 k_B ( T_{\parallel 2} -   T_{\perp 2} ),  \nonumber\\
\mathcal{C} &=& m n_2 v_2 \cos \xi_2 \left( \frac{k_B}{2 m}(T_{\parallel 2} + 4 T_{\perp 2} )  +\frac{B_2^2 \sin ^2\theta_2}{4 \pi  m n_2}+\frac{v_2^2}{2}\right).\nonumber
\end{eqnarray}

The anisotropic upstream version of these equations is obtained by replacing the right-hand side of each equation by the left-hand side, changing subscripts ``2'' to ``1'' and then setting $\xi_1=0$. Indeed, in case a shock propagates behind another one, the downstream of the first shock is eventually the upstream of the next one. A formalism accounting for an anisotropic upstream is therefore necessary since our model always leaves an anisotropic downstream (unless $\mathbf{B}_1=0$).

Even though the model can be solved, the algebra is extremely involved. The system is symbolically solved with \emph{Mathematica}. Its solutions are then numerically studied in \emph{MATLAB}. On occasions, the \emph{Mathematica} calculations give rise to the resolution of a polynomial of considerable length. In such cases, the polynomial is transferred to \emph{MATLAB} using the \emph{Mathematica} Notebook described in \cite{BretMM2010}.

It is useful to focus on the quantity
\begin{equation}\label{eq:T2}
T2 \equiv \tan\theta_2,
\end{equation}
as the system of equations above allows to deduce a polynomial equation for $T2$, easy to solve numerically. The general pattern of the resolution consists therefore in deriving such a polynomial and from its roots, to compute the other downstream quantities like $n_2$, in terms of the upstream parameters.

The following dimensionless variables are used throughout this work,
\begin{equation}\label{eq:dimless}
r = \frac{n_2}{n_1}, ~~~~~ \mathcal{M}_{A1} = \sqrt{\frac{m n_1 v_1^2}{B_1^2/4\pi}}, ~~~~~ \sigma = \frac{B_1^2/4\pi}{m n_1 v_1^2} = \frac{1}{\mathcal{M}_{A1}^2}.
\end{equation}
While the Alfv\'{e}n Mach number $\mathcal{M}_{A,i}$ is prominent in shock literature, the related $\sigma$ parameter is typically used in PIC simulations like \cite{Haggerty2022}.

In order to simplify the problem, in the present work we restrict to the case $T_1=0$, that is, the strong sonic shock case. This is why no sonic Mach number is defined above.

The upstream is therefore characterized by 4 variables: $n_1$, $\theta_1$, $B_1$ and $v_1$.

The downstream is characterized by 6 variables $n_2$, $\theta_2$, $B_2$, $v_2$, $\xi_2$ and $T_e$. The 6 equations (\ref{eq:conser1}-\ref{eq:conser6})  allow then to solve the problem.

We now outline the resolution of the conservation equations for Stage 1, Stage-2-firehose, and Stage-2-mirror.

\section{Study of Stage 1}\label{sec:stage1}
With $T_{1\parallel}=T_{1\perp}=0$, Eqs. (\ref{eq:CGLoblique1},\ref{eq:CGLoblique2}) for Stage 1 read,
\begin{eqnarray}\label{eq:CGLS1}
  T_{2\parallel} &=&   T_e\cos^2\theta_2,  \\
  T_{2\perp} &=&   \frac{1}{2}T_e\sin^2\theta_2. \nonumber
\end{eqnarray}

\subsection{Symmetries}
Although not immediately visible, the system (\ref{eq:conser1}-\ref{eq:conser6}) with prescriptions (\ref{eq:CGLS1}) has some symmetries.

It can be checked that all other things being equal, if the set of angles $(\theta_1,\theta_2,\xi_2)$ is a solution, then $(-\theta_1,-\theta_2,-\xi_2)$ is also a solution, while $(-\theta_1,+\theta_2,\pm \xi_2)$ is \emph{not}. This implies that we cannot ignore the negative $\theta_2$'s. We shall then restrict our exploration to $\theta_1 \in [0,\pi/2]$ and solve for $\theta_2,\xi_2 \in [-\pi/2,\pi/2]$.

\subsection{Resolution}
Resolving Stage 1 is then achieved through the following steps,
\begin{itemize}
  \item Eliminate $v_2$ everywhere by extracting its value from Eq. (\ref{eq:conser1}).
  \item Eliminate $B_2$ everywhere by extracting its value from Eq. (\ref{eq:conser2}).
  \item Use the resulting Eq. (\ref{eq:conser4}) to eliminate $T_e$.
  \item At this junction, we are left with $n_2$, $\theta_2$ and $\xi_2$ as unknowns. $\xi_2$ can be eliminated (defining $X2\equiv\tan\xi_2$). We finally obtain 2 equations for $r=n_2/n_1$ and $T2 = \tan\theta_2$.
\end{itemize}
The equation for $T2$ reads,
\begin{equation}\label{eq:EquaT2S1}
 (T2 \cos \theta_1-\sin \theta_1) \underbrace{\sum_{k=0}^9a_k T2^k}_{\equiv \Lambda} =0,
\end{equation}
with,
\begin{eqnarray}\label{eq:coeffs}
 a_0  &=& -128 \left(-2 \mathcal{M}_{A1}^2+\cos (2 \theta_1)+1\right)^2 \sin \theta_1, \nonumber\\
 a_1  &=&  128 \cos \theta_1 \left(4 \mathcal{M}_{A1}^4-2 \mathcal{M}_{A1}^2+\left(2-6 \mathcal{M}_{A1}^2\right) \cos (2 \theta_1)+\cos (4 \theta_1)+1\right),\nonumber\\
 a_2  &=&  -16 \left(16 \mathcal{M}_{A1}^4-16 \mathcal{M}_{A1}^2+\left(8-16 \mathcal{M}_{A1}^2\right) \cos (2 \theta_1)+\cos (4 \theta_1)+7\right) \sin \theta_1,\nonumber\\
 a_3  &=&  8 \cos \theta_1 \left(16 \mathcal{M}_{A1}^4-44 \mathcal{M}_{A1}^2+\left(48-68 \mathcal{M}_{A1}^2\right) \cos (2 \theta_1)+17 \cos (4 \theta_1)+31\right),\nonumber\\
 a_4  &=&  4 \left(-32 \mathcal{M}_{A1}^4+40 \cos (2 \theta_1) \mathcal{M}_{A1}^2+40 \mathcal{M}_{A1}^2+\cos (4 \theta_1)-1\right) \sin \theta_1,\nonumber\\
 a_5  &=&  4 \cos \theta_1 \left(-16 \left(2 \mathcal{M}_{A1}^4+\mathcal{M}_{A1}^2\right)+8 \left(7-2 \mathcal{M}_{A1}^2\right) \cos (2 \theta_1)+15 \cos (4 \theta_1)+41\right),\nonumber\\
 a_6  &=&  -2 \left(96 \mathcal{M}_{A1}^4-112 \mathcal{M}_{A1}^2+8 \left(3-14 \mathcal{M}_{A1}^2\right) \cos (2 \theta_1)+9 \cos (4 \theta_1)+15\right) \sin \theta_1,\nonumber\\
 a_7  &=&  2 \cos \theta_1 \left(32 \mathcal{M}_{A1}^4+16 \mathcal{M}_{A1}^2+8 \left(5-6 \mathcal{M}_{A1}^2\right) \cos (2 \theta_1)+15 \cos (4 \theta_1)+25\right),\nonumber\\
 a_8  &=&  16 \cos ^4\theta_1 \sin \theta_1,\nonumber\\
 a_9  &=&  16 \cos ^5\theta_1.
\end{eqnarray}
The equation for $r$ reads,
\begin{equation}\label{eq:rS1}
r=\frac{4 \mathcal{M}_{A1}^2 T2^3 (1+T2^2)}{\sum_{k=0}^5b_k T2^k},
\end{equation}
where,
\begin{eqnarray}
  b_0 &=& 8 \mathcal{M}_{A1}^2 \tan \theta_1-4 \sin (2 \theta_1), \nonumber\\
  b_1 &=& -8 \mathcal{M}_{A1}^2+6 \cos (2 \theta_1)+2, \nonumber\\
  b_2 &=& 0, \nonumber\\
  b_3 &=& 4 \mathcal{M}_{A1}^2+\cos (2 \theta_1)+3, \nonumber\\
  b_4 &=& 4 \mathcal{M}_{A1}^2 \tan \theta_1-2 \sin (2 \theta_1), \nonumber\\
  b_5 &=&  2 \cos ^2\theta_1.
\end{eqnarray}

Eq. (\ref{eq:EquaT2S1}) is a polynomial yielding various $T2$-branches as solutions. Scanning them, and using Eq. (\ref{eq:rS1}), allows to derive the density jump. Note that one value of $T2$ gives one single value of $r$.

\begin{figure}
  \centering
  \includegraphics[width=0.6\textwidth]{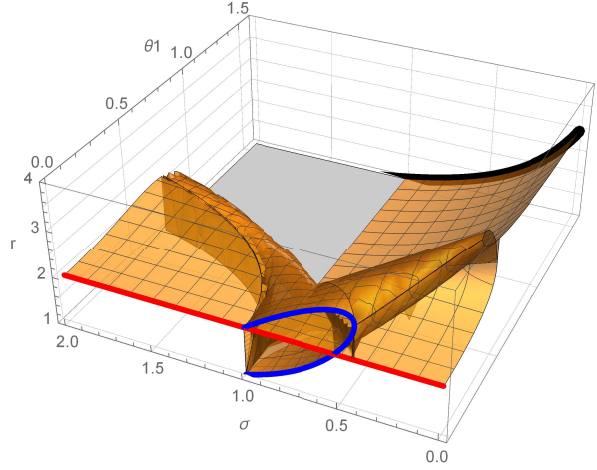}
  \caption{Values of the density ratio $r$ for Stage 1, with $\theta_2$ given by $\Lambda=0$ as defined by Eq. (\ref{eq:EquaT2S1}). The red curve was studied in \cite{BretJPP2018}. The blue one was studied in \cite{BretJPP2022}. The blue curve has $\theta_2>0$. The red one has $\theta_2 = 0$. The thick black curve at $\theta_1=\pi/2$ was studied in \cite{BretPoP2019}.}\label{fig:rS1}
\end{figure}

Eq. (\ref{eq:EquaT2S1}) clearly displays 2 main branches,
\begin{itemize}
  \item $T2 \cos \theta_1-\sin \theta_1 = 0$, that is, $\theta_2=\theta_1$. Inserting in Eq. (\ref{eq:rS1}) gives $r=1$. This is the continuity solution.
  \item $\Lambda=0$. The values of the density ratio $r$ so defined are represented on figure \ref{fig:rS1} in terms of ($\sigma,\theta_1$). For $\theta_1=0$, we recover the solutions found in \cite{BretJPP2018,BretJPP2022}. For $\theta_1=\pi/2$, we recover the solutions found in \cite{BretPoP2019}.
\end{itemize}

All these Stage 1 solutions do not make their way to the end state of the downstream since some are unstable. We need now to assess the stability of Stage 1.

\subsection{Stability of Stage 1}\label{sec:S1Stab}
From Eqs. (\ref{eq:firehose},\ref{eq:mirror}), we see that assessing the stability of Stage 1 requires computing its anisotropy $A_2$ and its $\beta_{2\parallel}$ parameter. The anisotropy for Stage 1 is straightforwardly given by Eqs. (\ref{eq:CGLS1}) as,
\begin{equation}\label{eq:A2S1}
A_2 = \frac{T_{2\perp}}{ T_{2\parallel}} =  \frac{1}{2}T_e\tan^2\theta_2.
\end{equation}
The $\beta_{\parallel 2}$ parameter is given by,
\begin{eqnarray}\label{eq:Beta2S1}
\beta_{\parallel 2}  = 2 \frac{\sec^2\theta_1 \left(2 \mathcal{M}_{A1}^2 (r-1)+r\right)-r \left(T2^2+1\right)}{r \left(T2^4+2\right)}.
\end{eqnarray}

Using Eqs. (\ref{eq:A2S1},\ref{eq:Beta2S1}) we can then numerically assess the firehose or mirror instability of Stage 1. Depending on the result, Stage 1 will be the end state of the downstream, or else it will migrate to Stage-2-firehose or Stage-2-mirror, on the corresponding instability thresholds.

\section{Study of Stage-2-firehose}\label{sec:S2fire}
In case Stage 1 is firehose unstable, it will migrate to the firehose stability threshold. In order to determine its properties, we need now to impose condition (\ref{eq:firehose}) to the system (\ref{eq:conser1}-\ref{eq:conser6}) instead of the temperatures prescriptions (\ref{eq:CGLS1}).

The resolution strategy is similar to that for Stage 1. Now $T2=\tan\theta_2$ is given solving,
\begin{equation}\label{eq:T2S2Fire}
   \sum_{k=0}^4a_k T2^k = 0,
\end{equation}
with,
\begin{eqnarray}\label{eq:coefsfire}
  a_0 &=& -32 \left(\mathcal{M}_{A1} \sin (2 \theta_1)-2 \mathcal{M}_{A1}^3 \tan \theta_1\right)^2,\nonumber\\
  a_1 &=& -10 \mathcal{M}_{A1}^2 \left[\left(20 \mathcal{M}_{A1}^2-2\right) \sin (2 \theta_1)-8 \left(2 \mathcal{M}_{A1}^2+1\right) \tan \theta_1 \mathcal{M}_{A1}^2-3 \sin (4 \theta_1)\right],\nonumber\\
  a_2 &=& \mathcal{M}_{A1}^2 \left[-32 \mathcal{M}_{A1}^4+8 \mathcal{M}_{A1}^2+24 \left(3 \mathcal{M}_{A1}^2-1\right) \cos (2 \theta_1)-15 \cos (4 \theta_1)-9\right],\nonumber\\
  a_3 &=& 0,\nonumber\\
  a_4 &=& -8 \mathcal{M}_{A1}^2 \cos ^4\theta_1 .
\end{eqnarray}
Then the density jump reads,
\begin{equation}\label{eq:rS2Fire}
  r = \frac{2 \mathcal{M}_{A1}^2 T2}{-\sin (2 \theta_1)+2 \mathcal{M}_{A1}^2 \tan \theta_1+T2 \cos ^2\theta_1}.
\end{equation}
Again, 1 value of $T2$ corresponds to one and only one value of $r$.

\begin{figure}
  \centering
  \includegraphics[width=0.48\textwidth]{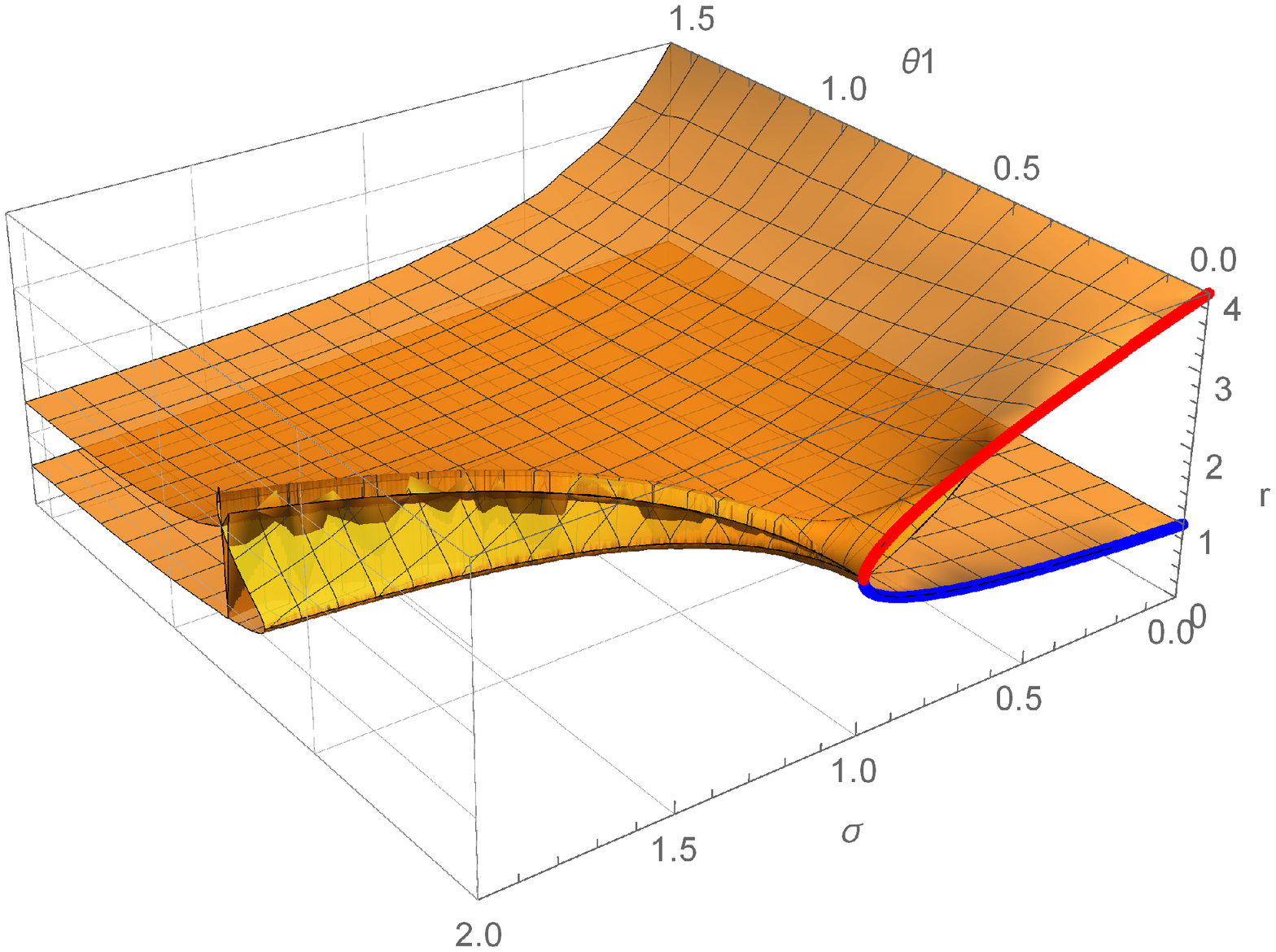} \includegraphics[width=0.48\textwidth]{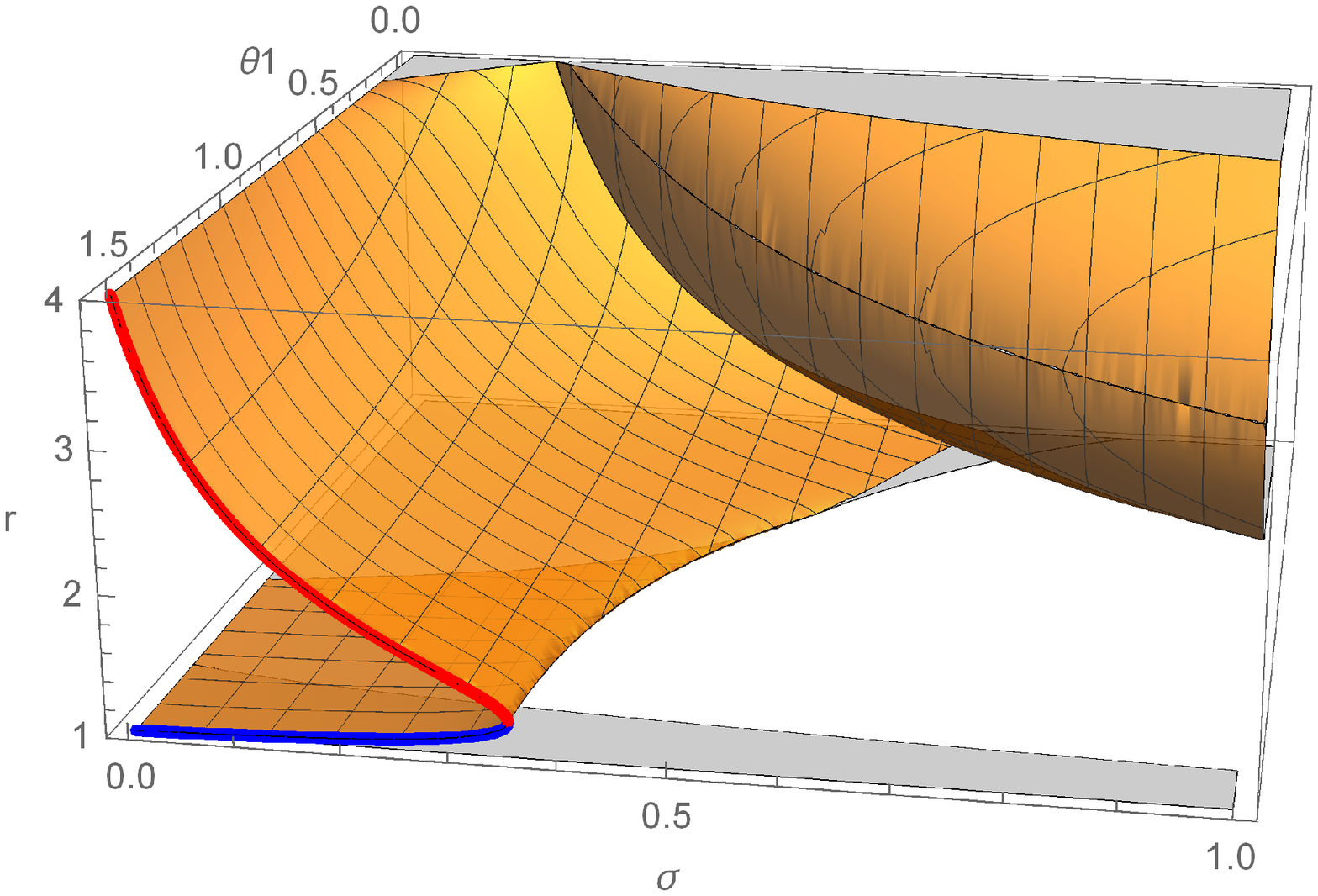}
  \caption{\emph{Left}: Values of $r$ for Stage-2-firehose, where $\theta_2$ is given by Eq. (\ref{eq:T2S2Fire}). The blue and red curves were studied in \cite{BretJPP2018}. The blue one has temperature anisotropy $A_2<0$ and is therefore nonphysical. \emph{Right}: Values of $r$ for Stage-2-mirror. The blue and red curves were studied in \cite{BretPoP2019}. The blue one has $A_2<0$. }\label{fig:rS2firemirror}
\end{figure}

The density jump so defined is plotted in figure \ref{fig:rS2firemirror}-left in terms of $(\sigma,\theta_1)$. For $\theta_1=0$, the red arc joining $(\sigma=0,r=4)$ to $(\sigma=1,r=2)$ fits exactly what was found in \cite{BretJPP2018}. In \cite{BretJPP2018}, we argued that the blue arc, joining $(\sigma=0,r=1)$ to $(\sigma=1,r=2)$, was not a shock solution since it reaches $r=1$ for $\sigma=0$. In fact, these blue solutions are discarded on an even simpler physical ground: they have $A_2 < 0$. For Stage-2-firehose, the anisotropy is no longer given by (\ref{eq:A2S1}) but by,
\begin{equation}\label{eq:A2Sefire}
A_2=1-\frac{r \left(T2^2+1\right) \cos ^2\theta_1}{2 \mathcal{M}_{A1}^2 (r-1)+r \sin ^2\theta_1}.
\end{equation}
When computing this quantify for the lower, blue arc, and indeed for the whole lower surface in figure \ref{fig:rS2firemirror}-left, $A_2 < 0$ is found. This property will be useful when putting Stages 1 and 2 together in section \ref{sec:together}.

\section{Study of Stage-2-mirror}\label{sec:S2mirror}
If Stage 1 is mirror unstable we need to impose relation (\ref{eq:mirror}) to the conservation equations. The quantity $T2$ is still solution of the polynomial equation Eq. (\ref{eq:T2S2Fire}), where the coefficients are now,
\begin{eqnarray}
  a_0 &=& 4 \left(\sin (2 \theta_1)-2 \mathcal{M}_{A1}^2 \tan \theta_1\right)^2 ,\nonumber\\
  a_1 &=& \frac{1}{4} \left[2 \left(74 \mathcal{M}_{A1}^2-17\right) \sin (2 \theta_1)-40 \left(2 \mathcal{M}_{A1}^2+1\right) \tan \theta_1 \mathcal{M}_{A1}^2-27 \sin (4 \theta_1)\right] ,\nonumber\\
  a_2 &=& 4 \mathcal{M}_{A1}^4+8 \sin ^2\theta_1 \mathcal{M}_{A1}^2+30 \cos ^4\theta_1-\cos ^2\theta_1 \left(30 \mathcal{M}_{A1}^2+19 \sin ^2\theta_1\right) ,\nonumber\\
  a_3 &=& -\left(1-2 \mathcal{M}_{A1}^2+\cos (2 \theta_1)\right) \sin (2 \theta_1) ,\nonumber\\
  a_4 &=& 15 \cos ^4\theta_1 .
\end{eqnarray}
Then the density jump becomes,
\begin{equation}\label{eq:rS2Mirror}
  r = \frac{2 \mathcal{M}_{A1}^2 T2}{-\sin (2 \theta_1)+2 \mathcal{M}_{A1}^2 \tan \theta_1+3 T2 \cos ^2\theta_1}.
\end{equation}
 The results are plotted in figure \ref{fig:rS2firemirror}-right in terms of $(\sigma,\theta_1)$. A pattern similar to that of Stage-2-firehose emerges here. When treating the $\theta_1=\pi/2$ problem in \cite{BretPoP2019}, we discarded the lower branch in blue at $\theta_1=\pi/2$, arguing it is not a shock solution since it reaches $r=1$ for $\sigma=0$. It turns out that this branch again has anisotropy $A_2 < 0$. For Stage-2-mirror, this quantity reads,
\begin{equation}\label{eq:A2S2mirror}
A_2 = -\frac{-4 \mathcal{M}_{A1}^2 (r-1)+r T2^2 \cos (2 \theta_1)+r \left(T2^2-2\right)}{4 \mathcal{M}_{A1}^2 (r-1)-r \left(2 T2^2+1\right) \cos (2 \theta_1)-2 r T2^2+r},
\end{equation}
and is found negative on the blue arc in figure \ref{fig:rS2firemirror}-right, as well as along the lower surface that extends from this arc.

\section{Putting Stages 1 and 2 together}\label{sec:together}
We finally come to the point where we can assemble Stages 1 and 2. This has been performed in \emph{MATLAB} according to the following algorithm,
\begin{enumerate}
  \item Solve $\Lambda=0$ in Eq. (\ref{eq:EquaT2S1}) for $T2$ in Stage 1, and record all the branches of the solutions.
  \item Then scan each Stage 1 branch. If a Stage 1 state is found stable, then this is the end state of the downstream.
  \item If a Stage 1 state is found firehose unstable, then switch to Stage-2-firehose, end state of the downstream.
  \item If a Stage 1 state is found mirror unstable, then switch to Stage-2-mirror, end state of the downstream.
\end{enumerate}

Steps (\emph{c}) and (\emph{d}) can be non-trivial when, for an unstable Stage 1 state $(\sigma,\theta_1)$, there are more than one Stage 2 states with the same $(\sigma,\theta_1)$. Some criteria are needed in order to select one Stage 2 state among the possible solutions. We apply the following ones,
\begin{enumerate}
  \item Discard Stage 2 states with $A_2<0$ since they represent nonphysical solutions to the equations.
  \item In case degeneracy persists, select the Stage 2 state which has $\theta_2$ closest to the unstable Stage 1.
\end{enumerate}

We now check how this method retrieves our previous result, before applying it to any intermediate angle $\theta_1$.

\begin{figure}
  \centering
  \includegraphics[width=0.32\textwidth]{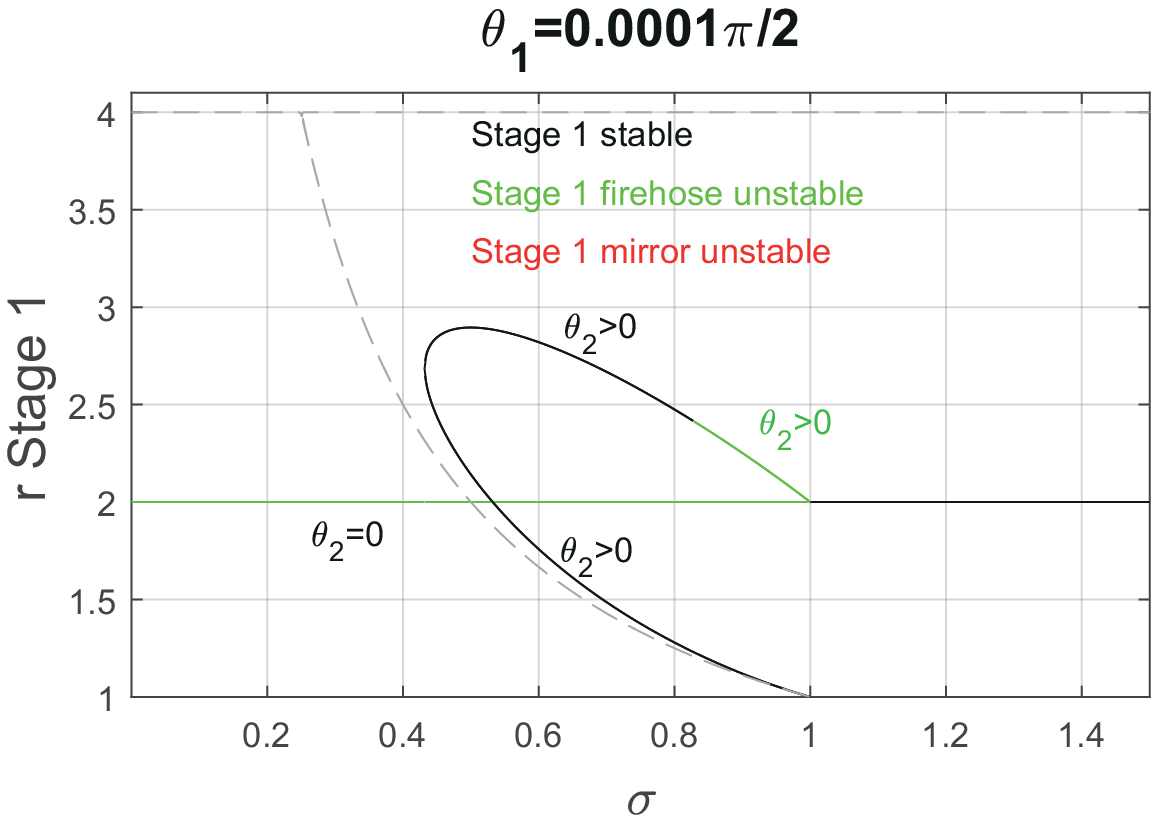} \includegraphics[width=0.32\textwidth]{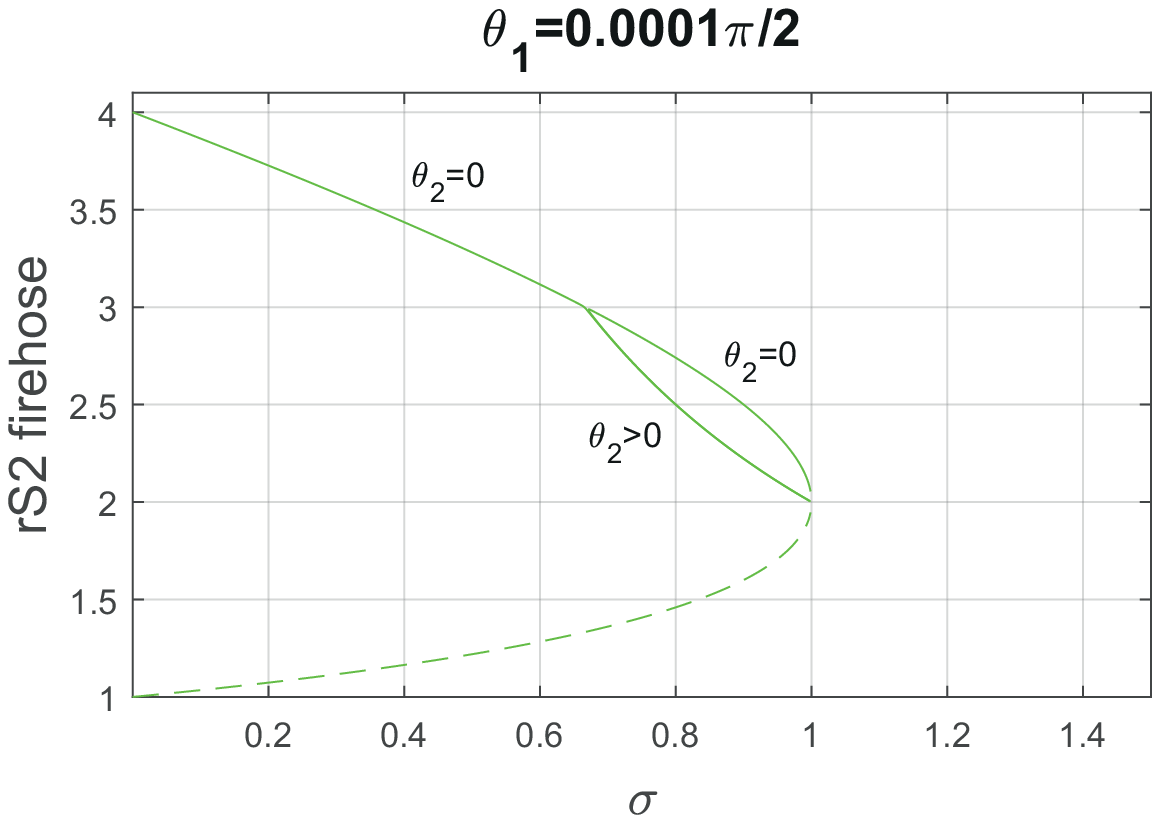} \includegraphics[width=0.32\textwidth]{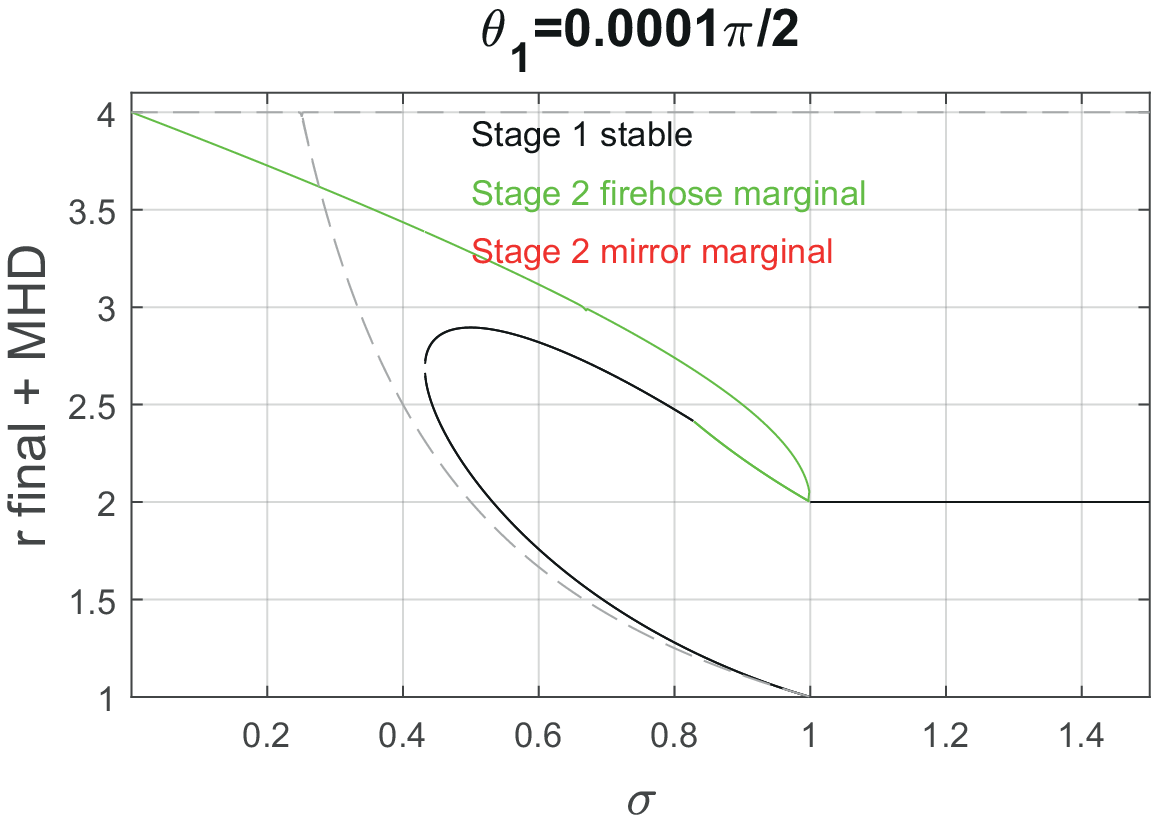}
  \caption{\emph{Left}: All solutions of Stage 1 for $\theta_1$ close to zero, color-coded according to their stability (none is mirror unstable). \emph{Center}: All Stage-2-firehose solutions. Dashed line indicate $A_2<0$ (nonphysical). \emph{Right}:  End result. The grey dashed lines show the MHD solutions.}\label{fig:theta0}
\end{figure}

\subsection{Case $\theta_1 \sim 0$}\label{sec:para}
The case $\theta_1 \sim 0$ is pictured on figure \ref{fig:theta0}. The left graph shows all Stage 1 solutions. Black means they are stable, green means they are firehose unstable. Red would mean mirror unstable, but for the selected $\theta_1$, there are no such cases. The solution $r=2$ has $\theta_1=\theta_2\sim 0$. It pertains to the parallel case which was studied in \cite{BretJPP2018}. The other solutions, which draw an open loop, pertain to the switch-on case studied in \cite{BretJPP2022}. They have $\theta_2 > 0$. These switch-on solutions are physical, namely, they have $A_2 > 0$ (see figure 4(a) of \cite{BretJPP2022}).

The center plot shows all solutions for Stage-2-firehose. We see that an unstable Stage 1 state with $\sigma = 0.9$, for example, can in principle go to 3 Stage-2-firehose states. Out of these 3, one has $A_2<0$, as indicated by the dashed line on the center plot. Among the 2 remaining options, the upper one has $\theta_2=0$ while the lower one has $\theta_2>0$. Therefore, choosing the Stage 2 state which has closest $\theta_2$ to the unstable Stage 1, leaves only 1 possible option.

The right plot shows the end result. We recover the result of the parallel case, with a marginal firehose jump going from $r=4$ to 2 for $0 < \sigma < 1$, and then Stage 1 stable with $r=2$ for $\sigma > 1$ \citep{BretJPP2018}. Also recovered are the 2 switch-on solutions found in \cite{BretJPP2022}, with a portion of the upper one being replaced by its Stage-2-firehose counterpart.

\begin{figure}
  \centering
  \includegraphics[width=0.32\textwidth]{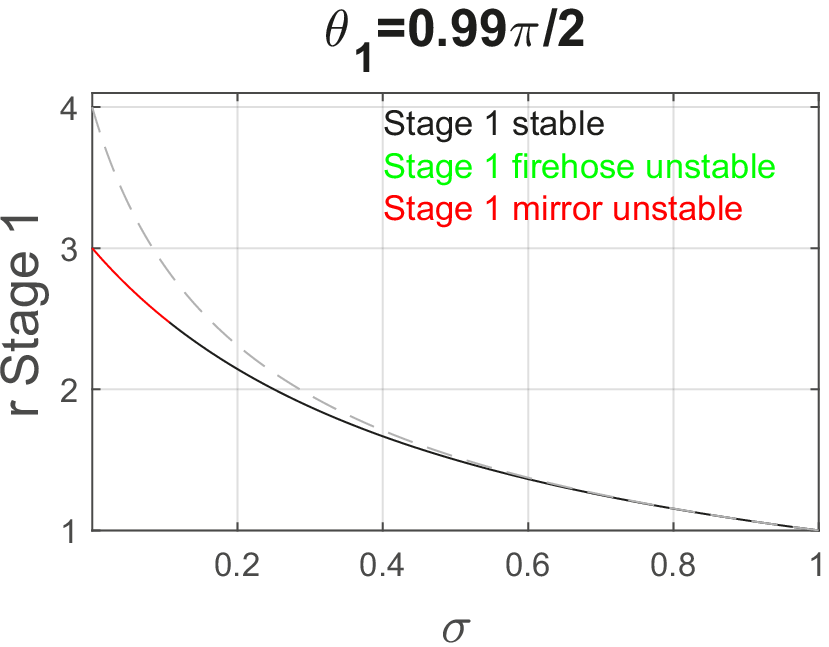} \includegraphics[width=0.32\textwidth]{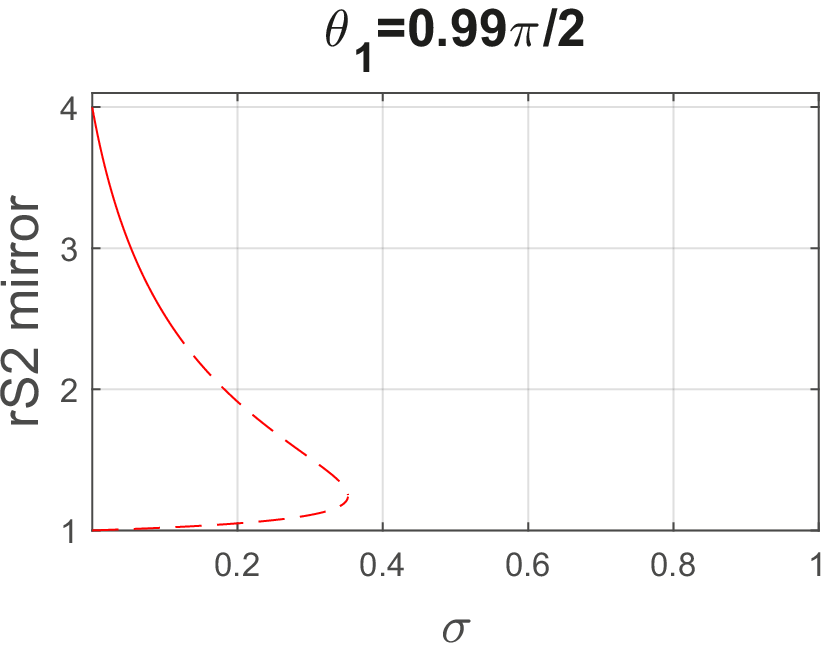} \includegraphics[width=0.32\textwidth]{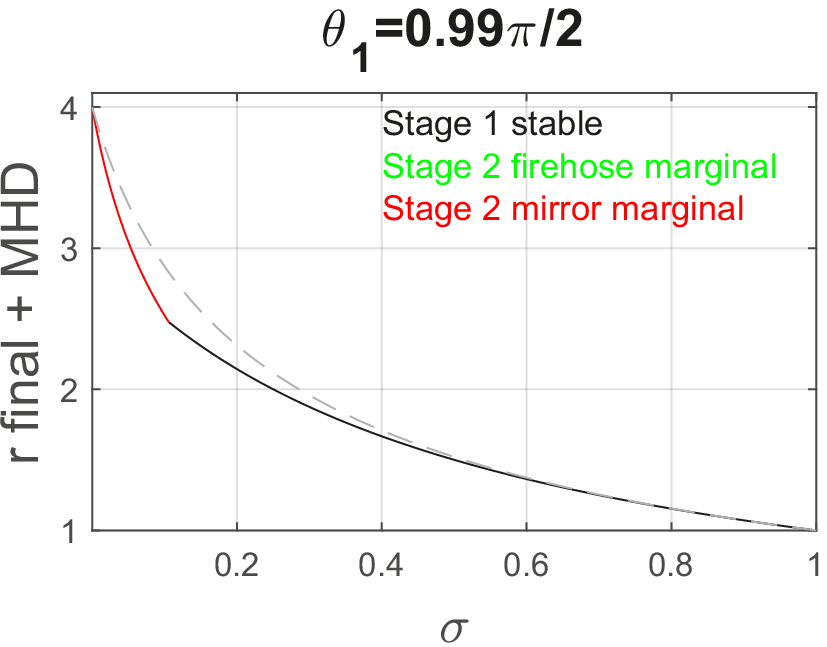}
  \caption{\emph{Left}: All solutions of Stage 1 for $\theta_1$ close to $\pi/2$, color-coded according to their stability (none is firehose unstable). \emph{Center}: All Stage-2-mirror solutions. Dashed line indicate $A_2<0$ (nonphysical). \emph{Right}:  End result. The grey dashed lines show the MHD solutions.}\label{fig:thetapi2}
\end{figure}

\subsection{Case $\theta_1 \sim \pi/2$}\label{sec:perp}
We here check the conformity of the present calculations with the results previously derived in \cite{BretPoP2019} for the perpendicular case.

Figure \ref{fig:thetapi2}-left shows Stage 1 solutions. There is but 1 branch solution, mirror unstable for $\sigma < \sigma_c$, where $\sigma_c=0.106$.

The center plot of figure \ref{fig:thetapi2} shows all of Stage-2-mirror branches. There is but one, with $A_2 < 0$ below $r \sim 2.47$, which is reached for $\sigma=\sigma_c'$. We checked numerically, up to the 13rd digit, that $\sigma_c = \sigma_c'$.

As a consequence, the right plot of figure \ref{fig:thetapi2}, which features the end result, has no gap. It fits exactly the result of \cite{BretPoP2019}. For $\sigma < \sigma_c$, Stage 1 is mirror unstable and the end state is Stage-2-mirror. Then for for $\sigma > \sigma_c$, Stage 1 is stable and gives the density jump of the end state.

\begin{figure}
  \centering
  \includegraphics[width=0.48\textwidth]{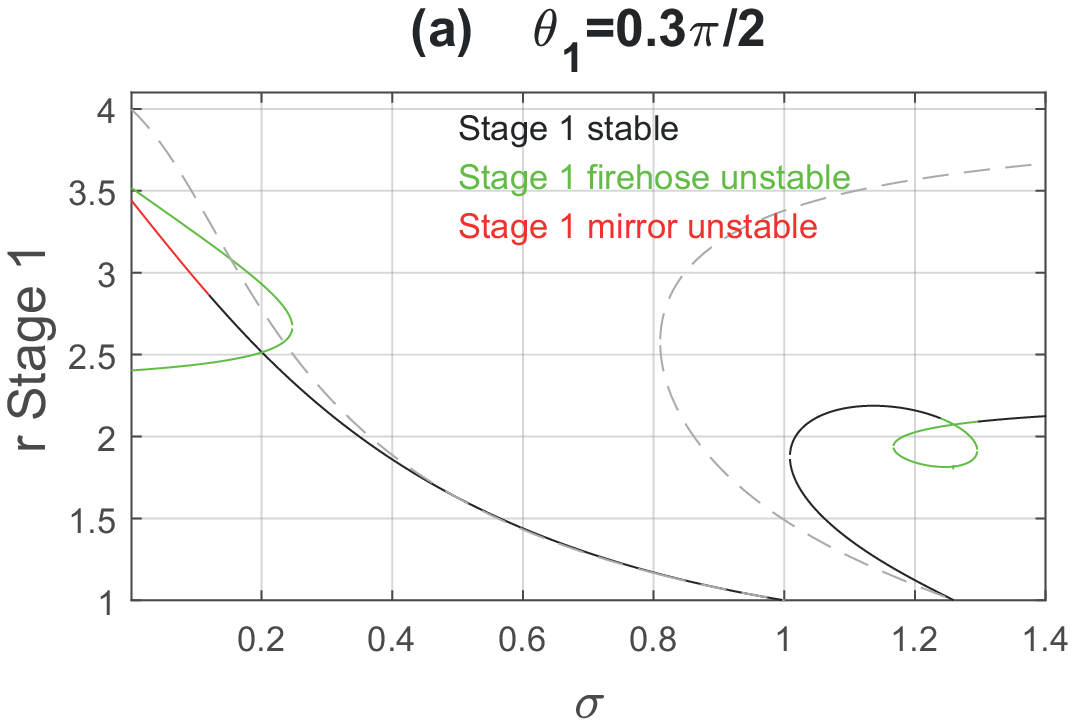} \includegraphics[width=0.48\textwidth]{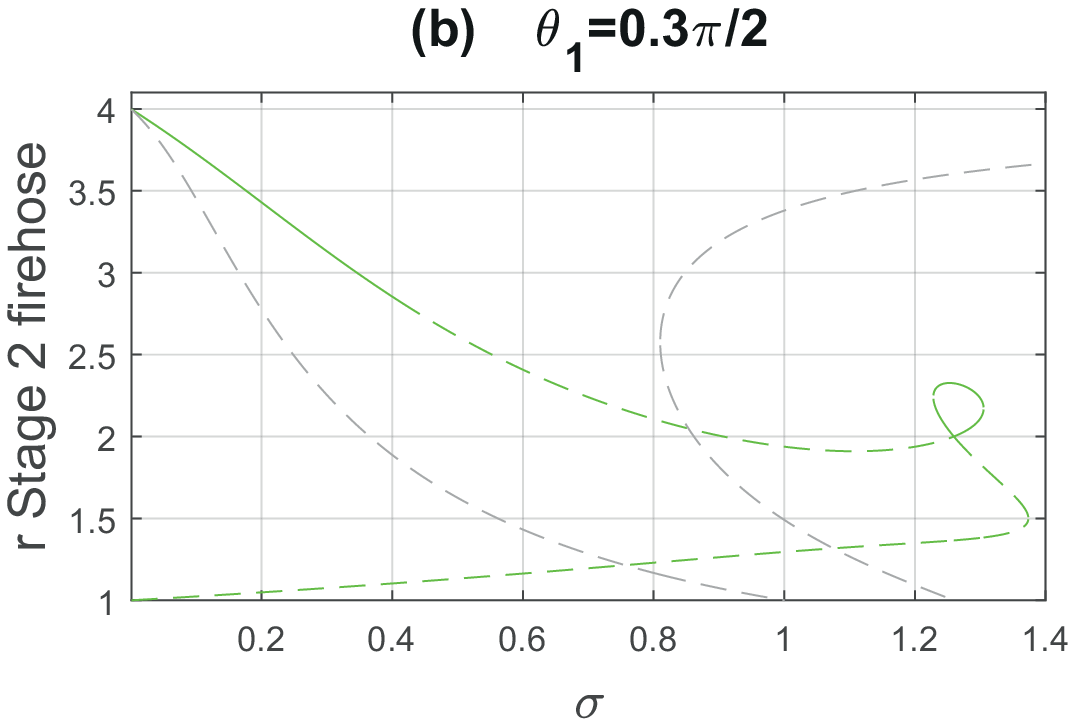} \\
  \includegraphics[width=0.48\textwidth]{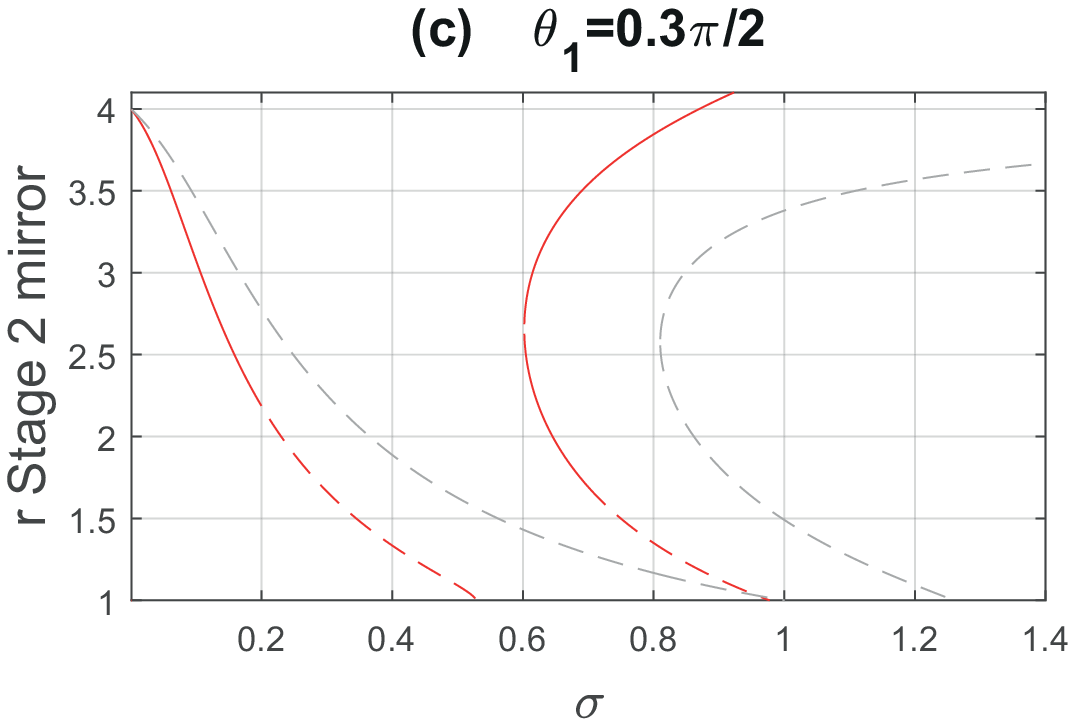} \includegraphics[width=0.48\textwidth]{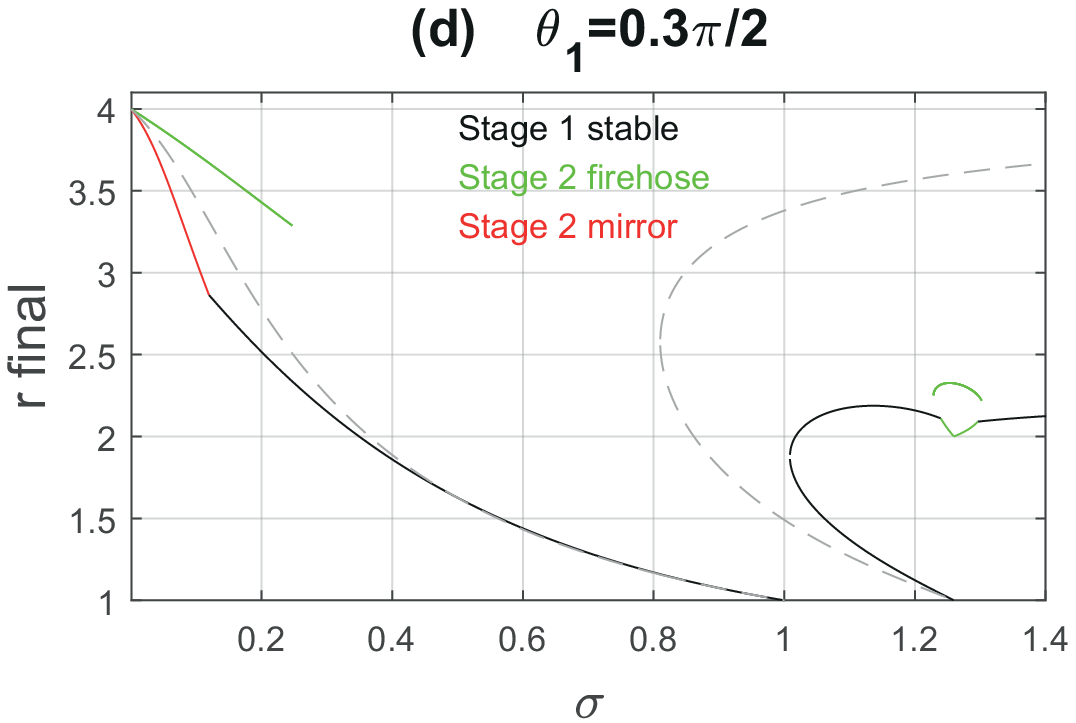}
  \caption{(a) All Stage 1 solutions for $\theta_1 = 0.3 \pi/2$, color-coded according to their stability. (b) All Stage-2-firehose solutions. Dashed line when $A_2 < 0$  (nonphysical). (c) All Stage-2-mirror solutions. Dashed line when $A_2 < 0$  (nonphysical). (d) End result. The grey dashed lines show the MHD solutions.}\label{fig:theta03}
\end{figure}

\subsection{General oblique case}\label{sec:oblique}
Figure \ref{fig:theta03} pictures the situation for an intermediate angle $\theta_1 = 0.3  \pi/2$. Figure \ref{fig:theta03} (a) shows all of Stage 1 solutions. Here, some are mirror unstable while others are firehose unstable. Looking at plots (c) and (b) we can see that there is always a Stage 2 solution when Stage 1 is unstable. For some values of $\sigma$, for example 0.1 or 1.25, there are various unstable Stage 1 solutions. As a consequence, figure \ref{fig:theta03} (d) displays various solutions for the end State corresponding to these $\sigma$.

Notice also how our solutions mimic the MHD solutions (dashed gray)  at low $\sigma$ and high $\sigma$.

\begin{figure}
  \centering
  \includegraphics[width=0.32\textwidth]{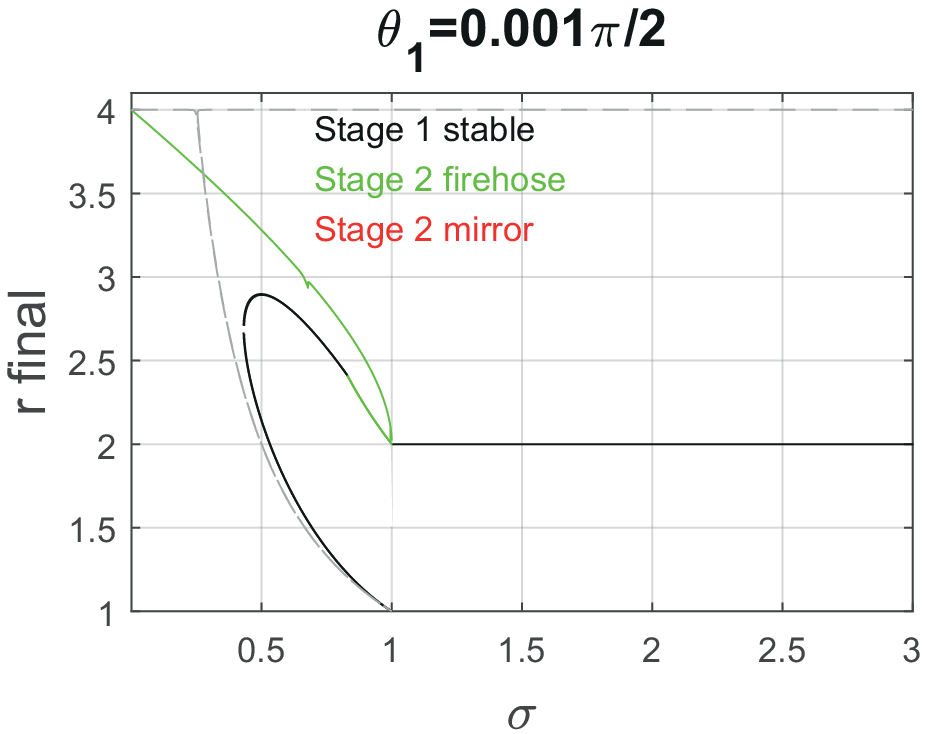}\includegraphics[width=0.32\textwidth]{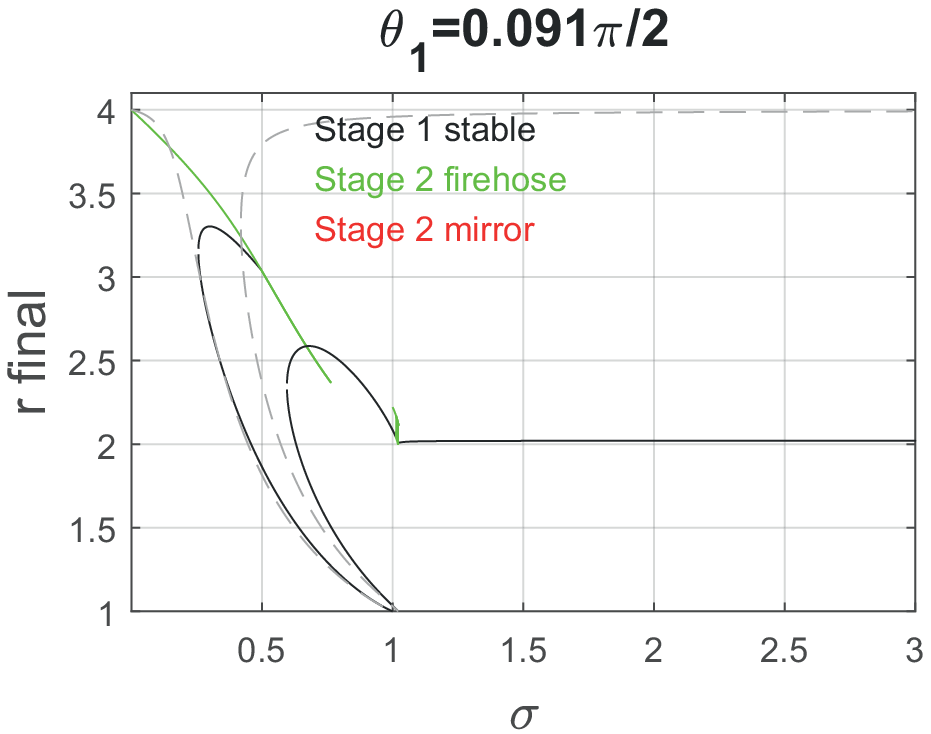}\includegraphics[width=0.32\textwidth]{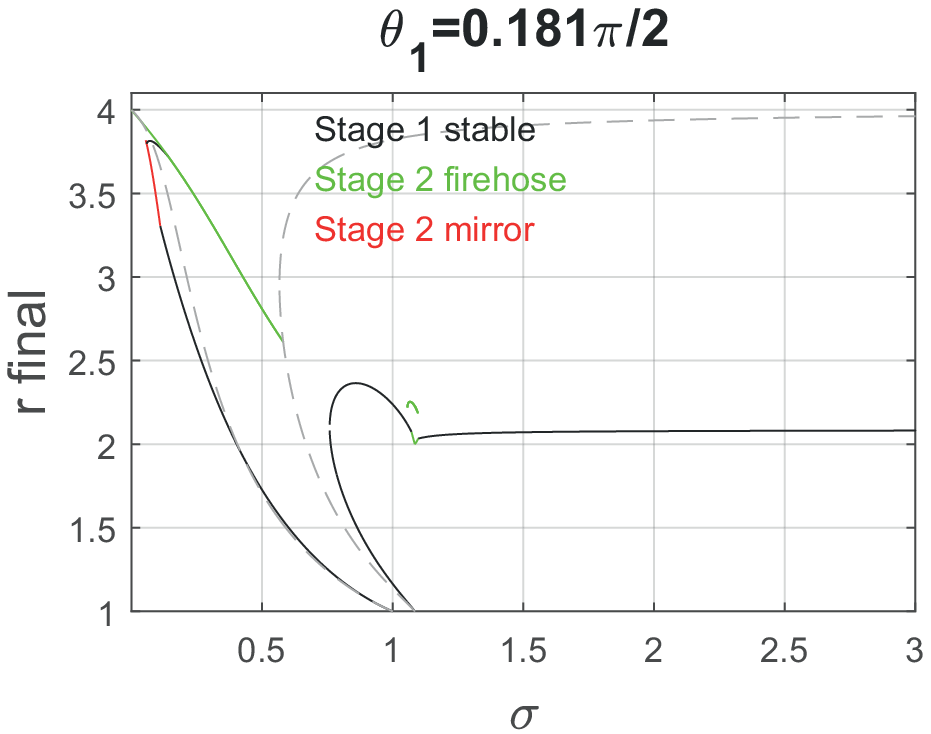}\\
    \includegraphics[width=0.32\textwidth]{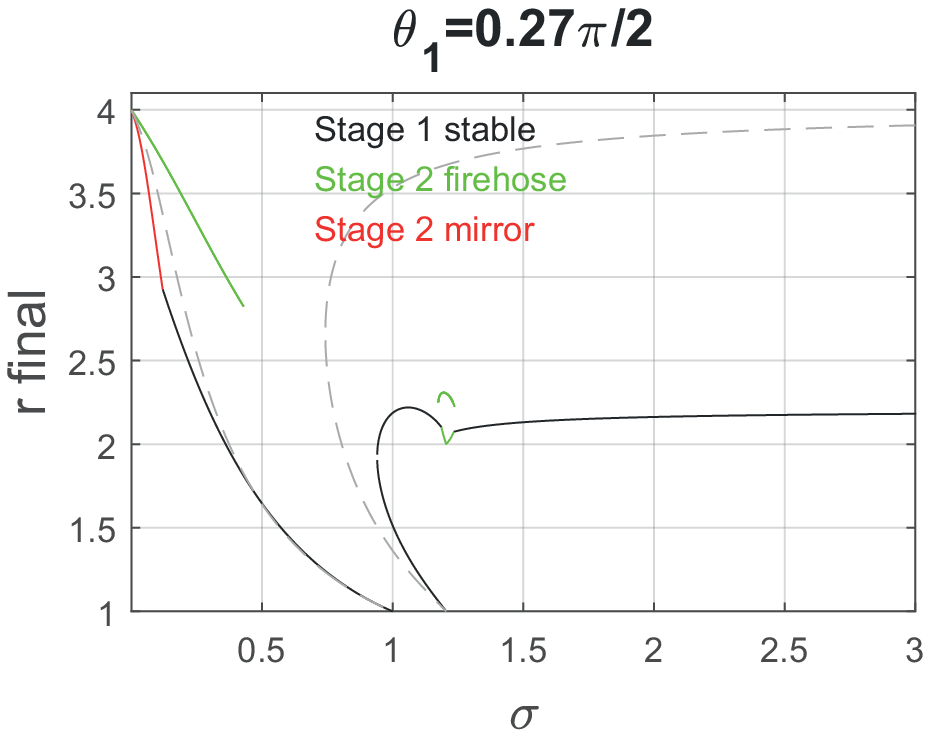}\includegraphics[width=0.32\textwidth]{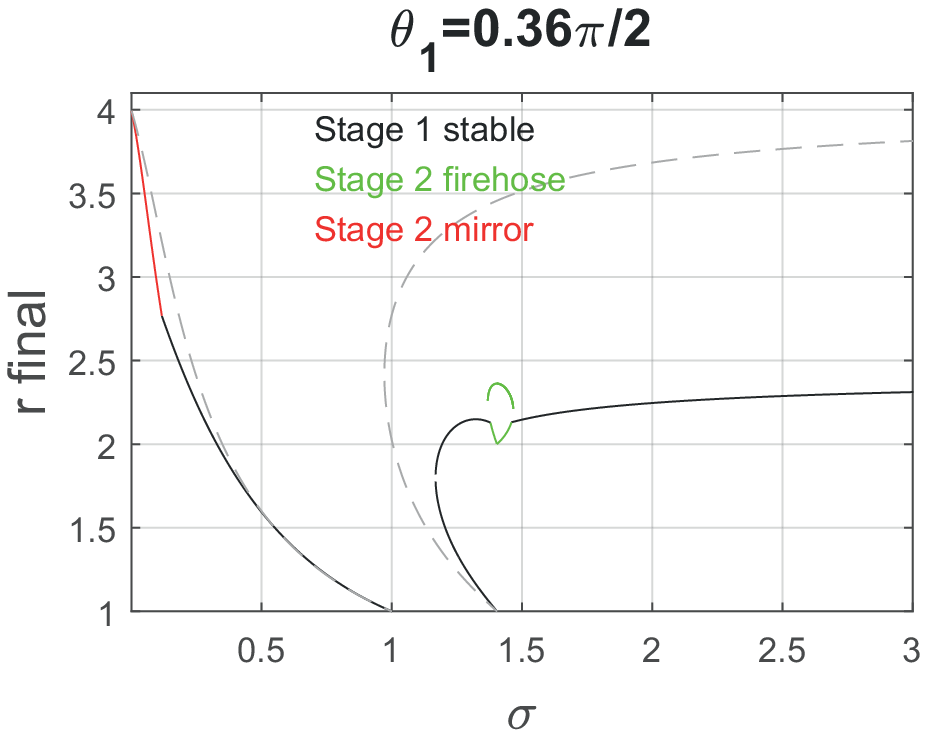}\includegraphics[width=0.32\textwidth]{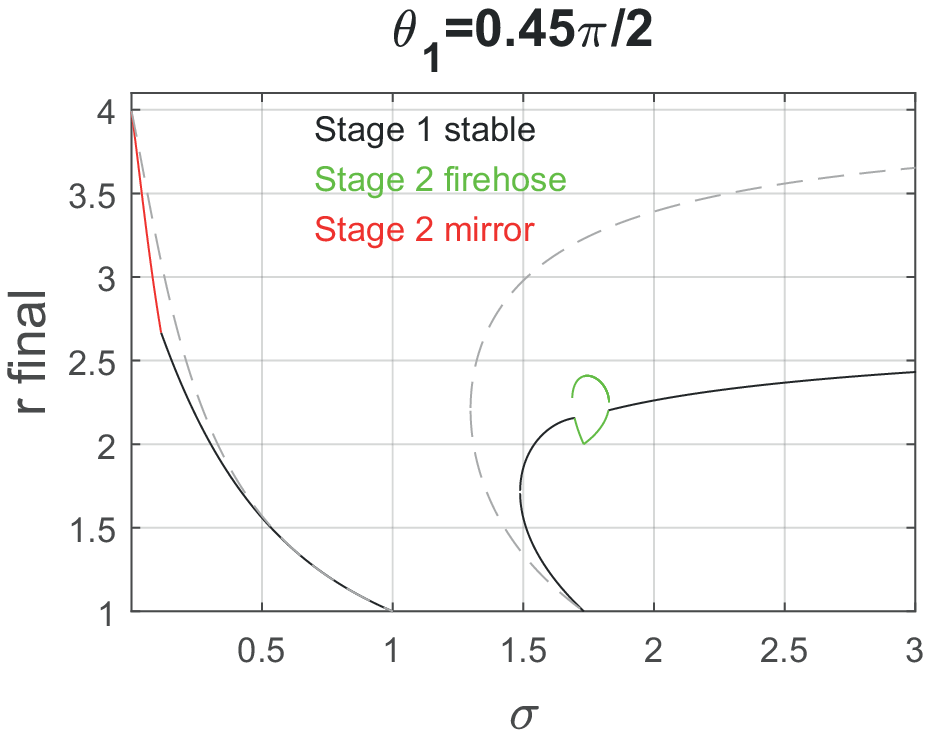}\\
  \includegraphics[width=0.32\textwidth]{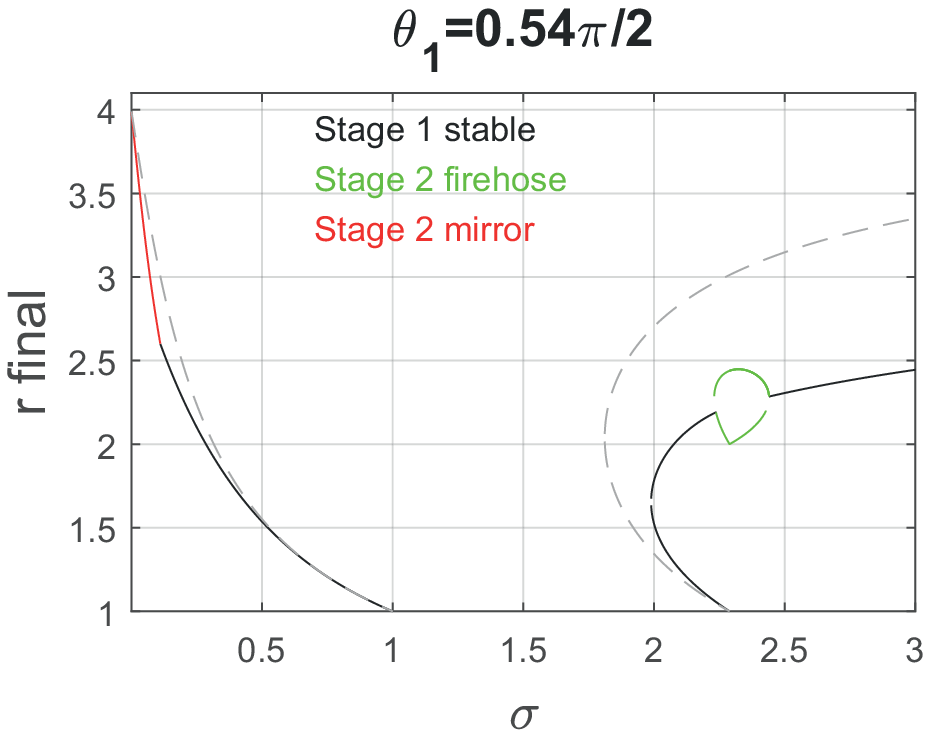}\includegraphics[width=0.32\textwidth]{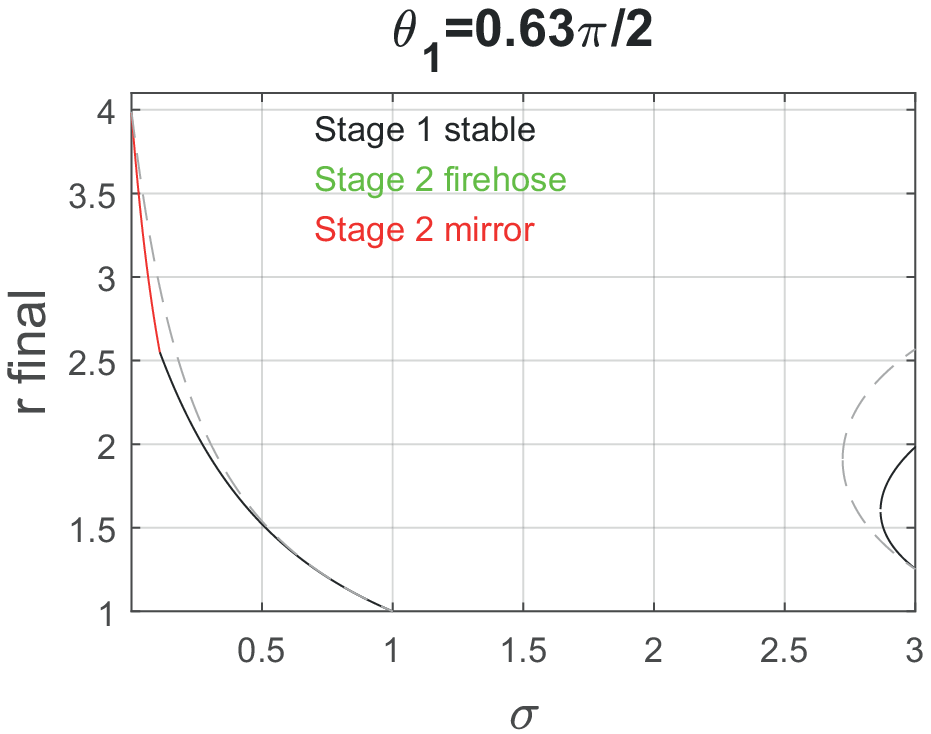}\includegraphics[width=0.32\textwidth]{cut08.eps}\\
  \includegraphics[width=0.32\textwidth]{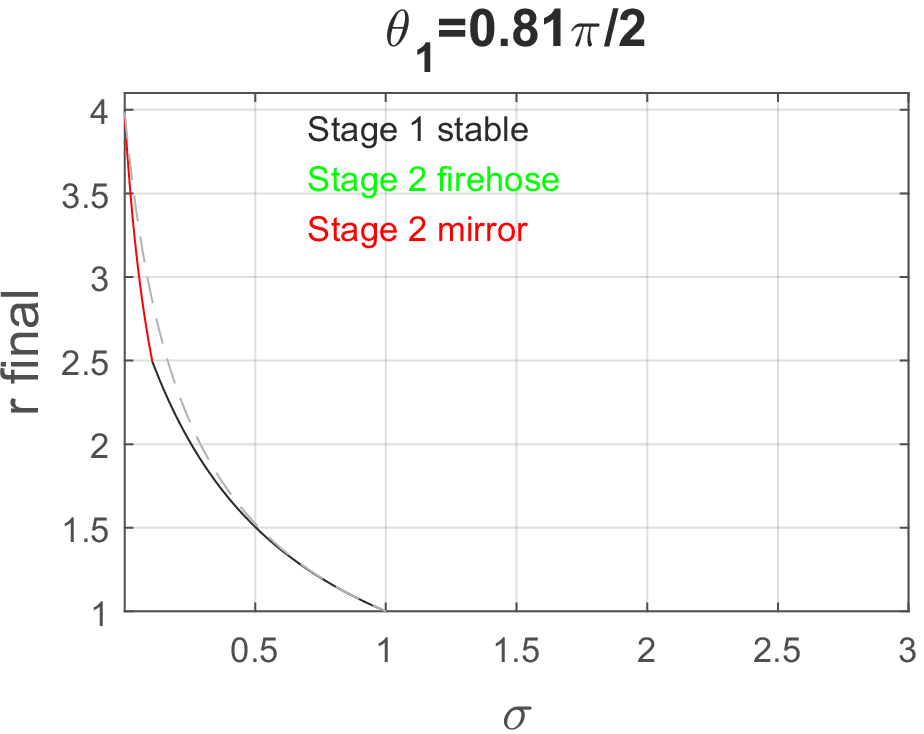}\includegraphics[width=0.32\textwidth]{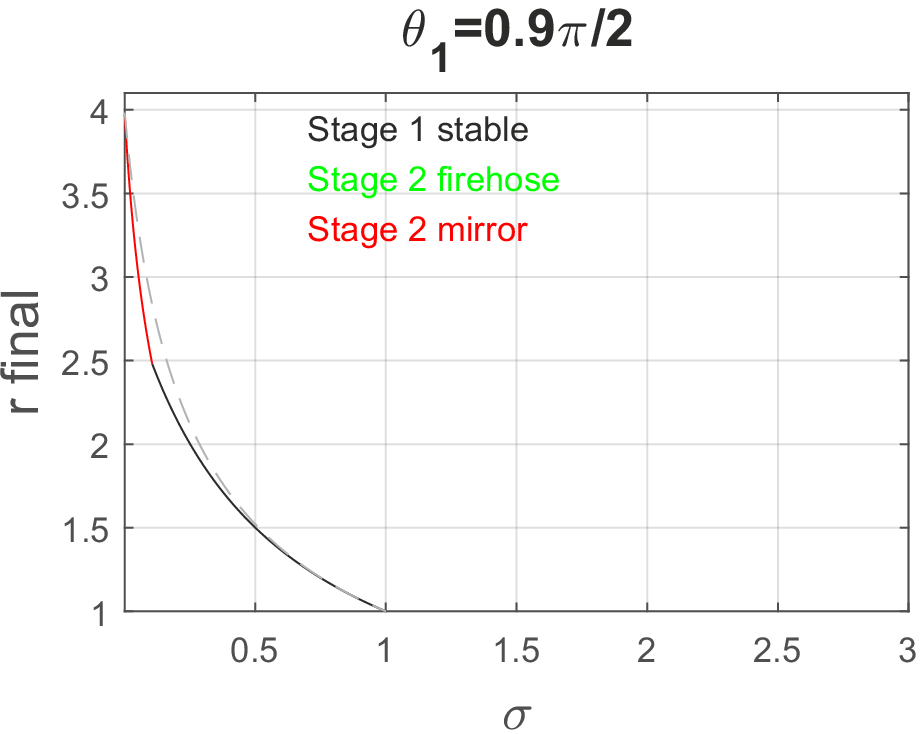}\includegraphics[width=0.32\textwidth]{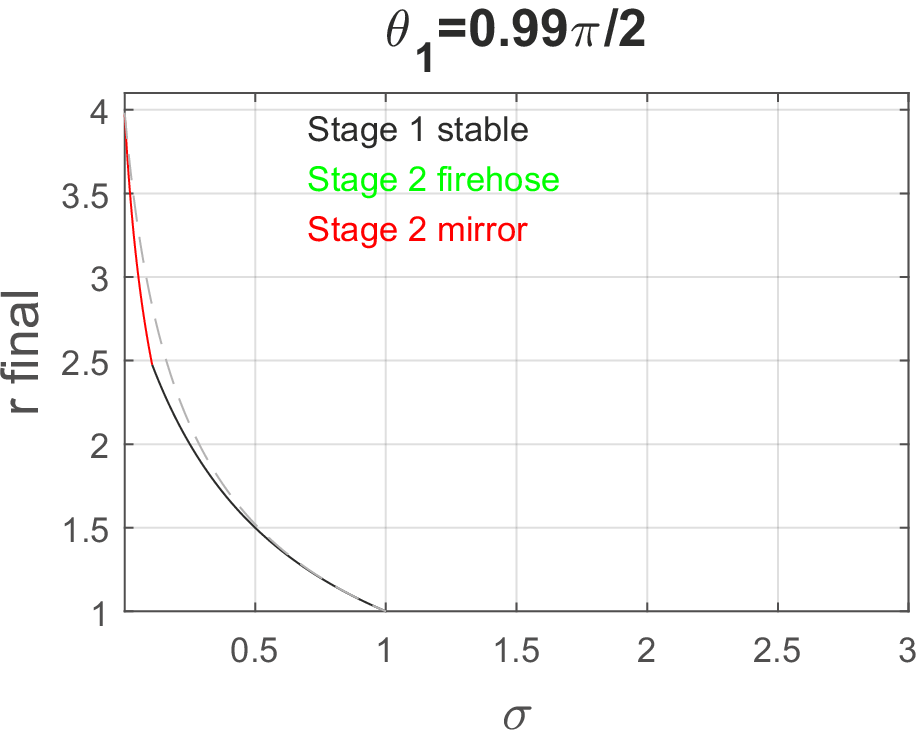}
  \caption{Similar to figure \ref{fig:theta03}(d), but for various values of $\theta_1 \in [0 , \pi/2]$. The grey dashed lines show the MHD solutions.}\label{fig:cuts}
\end{figure}

Finally, figure \ref{fig:cuts} presents a series of plots similar to figure \ref{fig:theta03}(d), for various values of $\theta_1 \in [0 , \pi/2]$.


\section{Conclusion}\label{sec:conclu}
In a series of recent articles, we elaborated a model of collisionless shocks. Having treated the parallel, the perpendicular and the switch-on cases \citep{BretJPP2018,BretPoP2019,BretJPP2022}, with our model for the parallel case being successfully tested against PIC simulations \citep{Haggerty2022}, we here treated the general oblique case.

MHD conservation equations for the general oblique case tend to be involved. MHD conservations equations for anisotropic temperatures are even more involved. And our model adds temperature prescriptions to these equations. As a result, its resolution is lengthy and requires extensive use of \emph{Mathematica}, to symbolically derive the key equations, and \emph{MATLAB}, to numerically solve them. The present work was devoted to the exposition of the mathematical solutions offered by our model. Their physics will be assessed in a forthcoming article.

In this respect, our model frequently offers various solutions for the same value of $\sigma$. Yet, Figs. \ref{fig:theta0} -left and -right show that such is also the case in MHD. This is also visible on Figs. \ref{fig:theta03} (a-d) and on most of Figs. \ref{fig:cuts}. In MHD, the solution selected depends on its physical relevance (see second to last paragraph of the introduction), or on the initial conditions of the shock formation like, for example, which initial states of a Riemann problem it is supposed to connect. In general, this second issue, namely connecting 2 different states, requires a succession of shocks rather than 1 single shock (see for example \cite{Ryu1995}). In our model, the choice of the solution when various are offered, will most probably depend on the same factors. This topic will be addressed in a forthcoming paper.

Even though only the case of a cold upstream has been solved here, the formalism allows in principle for an anisotropic upstream.

Although we treated the field obliquity as an arbitrary parameter, this study remains limited in various ways,
\begin{enumerate}
  \item A pair plasma is considered.
  \item Velocities are non-relativistic.
  \item The upstream pressure is assumed zero.
  \item The shock is coplanar, namely, upstream and downstream fields and velocities share a common plane.
\end{enumerate}

Regarding limitation ($a$), PIC simulations could be used to test the relevance of our model to electron/ion plasmas, provided the $\sigma$ parameter in  Eq. (\ref{eq:dimless}) is defined using the ion mass.

Tackling the other limitations altogether is clearly out of reach. At any rate, further testing of our model is envisioned through PIC simulations or comparison with \emph{in situ} measurements at interplanetary shocks, by spacecrafts like Advanced Composition Explorer, Wind or the Parker Solar Probe (see for example \cite{Fraschetti2022}).

As evidenced in figure \ref{fig:cuts}, deviations from MHD are more pronounced for quasi-parallel shocks and $\sigma > 1$. This is therefore the domain where our model should preferably be compared with PIC simulations or \emph{in situ} measurements.

\section*{Acknowledgments}
Thanks are due to Roberto Piriz for enriching discussions.

\section*{Funding}
A.B. acknowledges support by the Ministerio de Econom\'{i}a  y Competitividad of Spain (Grant No. PID2021-125550OB-I00).
R.N. acknowledges support from the NSF Grant No. AST-1816420. R.N. and A.B. thank the Black Hole Initiative at Harvard University for support. The BHI is funded by grants from the John Templeton Foundation and the Gordon and Betty Moore Foundation.

\section*{Declaration of Interests}
The authors report no conflict of interest.



\begin{thebibliography}{2}
\expandafter\ifx\csname natexlab\endcsname\relax\def\natexlab#1{#1}\fi

\bibitem[{Bret} \& {Narayan}(2018)]{BretJPP2018}
{\sc {Bret}, Antoine \& {Narayan}, Ramesh} 2018 {Density jump as a function of
  magnetic field strength for parallel collisionless shocks in pair plasmas}.
  {\em Journal of Plasma Physics\/} {\bf 84}~(6), 905840604.

\bibitem[{Haggerty} {\em et~al.\/}(2022){Haggerty}, {Bret} \&
  {Caprioli}]{Haggerty2022}
{\sc {Haggerty}, Colby~C., {Bret}, Antoine \& {Caprioli}, Damiano} 2022
  {Kinetic simulations of strongly magnetized parallel shocks: deviations from
  MHD jump conditions}. {\em Monthly Notices of the Royal Astronomical
  Society\/} {\bf 509}~(2), 2084--2090.

\end{thebibliography}


\begin{thebibliography}{35}
\expandafter\ifx\csname natexlab\endcsname\relax\def\natexlab#1{#1}\fi

\bibitem[Bale {\em et~al.\/}(2009)Bale, Kasper, Howes, Quataert, Salem \&
  Sundkvist]{BalePRL2009}
{\sc Bale, S.~D., Kasper, J.~C., Howes, G.~G., Quataert, E., Salem, C. \&
  Sundkvist, D.} 2009 Magnetic fluctuation power near proton temperature
  anisotropy instability thresholds in the solar wind. {\em Phys. Rev. Lett.\/}
  {\bf 103}, 211101.

\bibitem[Berezhko \& Ellison(1999)]{Berezhko1999}
{\sc Berezhko, E.~G. \& Ellison, Donald~C.} 1999 A simple model of nonlinear
  diffusive shock acceleration. {\em The Astrophysical Journal\/} {\bf
  526}~(1), 385--399.

\bibitem[{Bret}(2010)]{BretMM2010}
{\sc {Bret}, A.} 2010 Transferring a symbolic polynomial expression from
  \emph{Mathematica} to \emph{Matlab}. {\em ArXiv:1002.4725\/} .

\bibitem[{Bret}(2020)]{BretApJ2020}
{\sc {Bret}, Antoine} 2020 {Can We Trust MHD Jump Conditions for Collisionless
  Shocks?} {\em \apj\/} {\bf 900}~(2), 111.

\bibitem[{Bret} \& {Narayan}(2018)]{BretJPP2018}
{\sc {Bret}, Antoine \& {Narayan}, Ramesh} 2018 {Density jump as a function of
  magnetic field strength for parallel collisionless shocks in pair plasmas}.
  {\em Journal of Plasma Physics\/} {\bf 84}~(6), 905840604.

\bibitem[{Bret} \& {Narayan}(2019)]{BretPoP2019}
{\sc {Bret}, A. \& {Narayan}, R.} 2019 {Density jump as a function of magnetic
  field for collisionless shocks in pair plasmas: The perpendicular case}. {\em
  Physics of Plasmas\/} {\bf 26}~(6), 062108.

\bibitem[{Bret} \& {Narayan}(2020)]{BretLPB2020}
{\sc {Bret}, Antoine \& {Narayan}, Ramesh} 2020 {Density jump for parallel and
  perpendicular collisionless shocks}. {\em Laser and Particle Beams\/} {\bf
  38}~(2), 114--120.

\bibitem[{Bret} \& {Narayan}(2022)]{BretJPP2022}
{\sc {Bret}, Antoine \& {Narayan}, Ramesh} 2022 {Density jump as a function of
  magnetic field for switch-on collisionless shocks in pair plasmas}. {\em
  Journal of Plasma Physics\/} {\bf 88}~(3), 905880320.

\bibitem[Chew {\em et~al.\/}(1956)Chew, Goldberger \& Low]{CGL1956}
{\sc Chew, G.~F., Goldberger, M.~L. \& Low, F.~E.} 1956 The boltzmann equation
  and the one-fluid hydromagnetic equations in the absence of particle
  collisions. {\em Proceedings of the Royal Society of London A: Mathematical,
  Physical and Engineering Sciences\/} {\bf 236}~(1204), 112--118.

\bibitem[{David} {\em et~al.\/}(2022){David}, {Fraschetti}, {Giacalone},
  {Wimmer-Schweingruber}, {Berger} \& {Lario}]{Fraschetti2022}
{\sc {David}, Liam, {Fraschetti}, Federico, {Giacalone}, Joe,
  {Wimmer-Schweingruber}, Robert~F., {Berger}, Lars \& {Lario}, David} 2022 {In
  Situ Measurement of the Energy Fraction in Suprathermal and Energetic
  Particles at ACE, Wind, and PSP Interplanetary Shocks}. {\em \apj\/} {\bf
  928}~(1), 66.

\bibitem[{Delmont} \& {Keppens}(2011)]{Delmont2011}
{\sc {Delmont}, P. \& {Keppens}, R.} 2011 {Parameter regimes for slow,
  intermediate and fast MHD shocks}. {\em Journal of Plasma Physics\/} {\bf
  77}~(2), 207--229.

\bibitem[{Erkaev} {\em et~al.\/}(2000){Erkaev}, {Vogl} \&
  {Biernat}]{Erkaev2000}
{\sc {Erkaev}, N.~V., {Vogl}, D.~F. \& {Biernat}, H.~K.} 2000 {Solution for
  jump conditions at fast shocks in an anisotropic magnetized plasma}. {\em
  Journal of Plasma Physics\/} {\bf 64}, 561--578.

\bibitem[{Falle} \& {Komissarov}(1997)]{Komissarov1997}
{\sc {Falle}, S.~A.~E.~G. \& {Komissarov}, S.~S.} 1997 {On the Existence of
  Intermediate Shocks}. In {\em Computational Astrophysics; 12th Kingston
  Meeting on Theoretical Astrophysics\/} (ed. D.~A. {Clarke} \& M.~J. {West}),
  {\em Astronomical Society of the Pacific Conference Series\/}, vol.~12,
  p.~66.

\bibitem[{Feldman} {\em et~al.\/}(1982){Feldman}, {Bame}, {Gary}, {Gosling},
  {McComas}, {Thomsen}, {Paschmann}, {Sckopke}, {Hoppe} \&
  {Russell}]{Feldman1982}
{\sc {Feldman}, W.~C., {Bame}, S.~J., {Gary}, S.~P., {Gosling}, J.~T.,
  {McComas}, D., {Thomsen}, M.~F., {Paschmann}, G., {Sckopke}, N., {Hoppe},
  M.~M. \& {Russell}, C.~T.} 1982 {Electron Heating Within the Earth's Bow
  Shock}. {\em Physical Review Letters\/} {\bf 49}~(3), 199--201.

\bibitem[Gary(1993)]{Gary1993}
{\sc Gary, S.~Peter} 1993 {\em Theory of Space Plasma Microinstabilities\/}.
  Cambridge University Press.

\bibitem[{Gary} \& {Karimabadi}(2009)]{Gary2009}
{\sc {Gary}, S.~P. \& {Karimabadi}, H.} 2009 {Fluctuations in electron-positron
  plasmas: Linear theory and implications for turbulence}. {\em Physics of
  Plasmas\/} {\bf 16}~(4), 042104.

\bibitem[{G{\'e}not}(2009)]{Genot2009}
{\sc {G{\'e}not}, V.} 2009 {Analytical solutions for anisotropic MHD shocks}.
  {\em Astrophysics and Space Sciences Transactions\/} {\bf 5}~(1), 31--34.

\bibitem[Goedbloed {\em et~al.\/}(2010)Goedbloed, Keppens \&
  Poedts]{Goedbloed2010}
{\sc Goedbloed, J.P., Keppens, R. \& Poedts, S.} 2010 {\em Advanced
  Magnetohydrodynamics: With Applications to Laboratory and Astrophysical
  Plasmas\/}. Cambridge University Press.

\bibitem[{Goedbloed}(2008)]{Goedbloed2008}
{\sc {Goedbloed}, J.~P.} 2008 {Time reversal duality of magnetohydrodynamic
  shocks}. {\em Physics of Plasmas\/} {\bf 15}~(6), 062101.

\bibitem[{Guo} {\em et~al.\/}(2017){Guo}, {Sironi} \& {Narayan}]{Guo2017}
{\sc {Guo}, X., {Sironi}, L. \& {Narayan}, R.} 2017 {Electron Heating in
  Low-Mach-number Perpendicular Shocks. I. Heating Mechanism}. {\em \apj\/}
  {\bf 851}, 134.

\bibitem[{Guo} {\em et~al.\/}(2018){Guo}, {Sironi} \& {Narayan}]{Guo2018}
{\sc {Guo}, X., {Sironi}, L. \& {Narayan}, R.} 2018 {Electron Heating in Low
  Mach Number Perpendicular Shocks. II. Dependence on the Pre-shock
  Conditions}. {\em \apj\/} {\bf 858}, 95.

\bibitem[Gurnett \& Bhattacharjee(2005)]{gurnett2005}
{\sc Gurnett, D.A. \& Bhattacharjee, A.} 2005 {\em Introduction to Plasma
  Physics: With Space and Laboratory Applications\/}. Cambridge University
  Press.

\bibitem[{Haggerty} {\em et~al.\/}(2022){Haggerty}, {Bret} \&
  {Caprioli}]{Haggerty2022}
{\sc {Haggerty}, Colby~C., {Bret}, Antoine \& {Caprioli}, Damiano} 2022
  {Kinetic simulations of strongly magnetized parallel shocks: deviations from
  MHD jump conditions}. {\em Monthly Notices of the Royal Astronomical
  Society\/} {\bf 509}~(2), 2084--2090.

\bibitem[{Hasegawa}(1975)]{Hasegawa1975}
{\sc {Hasegawa}, A.} 1975 {Plasma instabilities and nonlinear effects}. {\em
  Springer Verlag Springer Series on Physics Chemistry Space\/} {\bf 8}.

\bibitem[{Hudson}(1970)]{Hudson1970}
{\sc {Hudson}, P.~D.} 1970 {Discontinuities in an anisotropic plasma and their
  identification in the solar wind}. {\em Planetary and Space Science\/} {\bf
  18}~(11), 1611--1622.

\bibitem[{Kennel} {\em et~al.\/}(1990){Kennel}, {Blandford} \&
  {Wu}]{Kennel1990}
{\sc {Kennel}, C.~F., {Blandford}, R.~D. \& {Wu}, C.~C.} 1990 {Structure and
  evolution of small-amplitude intermediate shock waves}. {\em Physics of
  Fluids B\/} {\bf 2}~(2), 253--269.

\bibitem[Kulsrud(2005)]{Kulsrud2005}
{\sc Kulsrud, Russell~M} 2005 {\em Plasma physics for astrophysics\/}.
  Princeton, NJ: Princeton Univ. Press.

\bibitem[Landau \& Lifshitz(1981)]{LandauKinetic}
{\sc Landau, L.D. \& Lifshitz, E.M.} 1981 {\em Course of Theoretical Physics,
  Physical Kinetics\/}, , vol.~10. Elsevier, Oxford.

\bibitem[Maruca {\em et~al.\/}(2011)Maruca, Kasper \& Bale]{MarucaPRL2011}
{\sc Maruca, B.~A., Kasper, J.~C. \& Bale, S.~D.} 2011 What are the relative
  roles of heating and cooling in generating solar wind temperature
  anisotropies? {\em Phys. Rev. Lett.\/} {\bf 107}, 201101.

\bibitem[{Ryu} \& {Jones}(1995)]{Ryu1995}
{\sc {Ryu}, Dongsu \& {Jones}, T.~W.} 1995 {Numerical Magnetohydrodynamics in
  Astrophysics: Algorithm and Tests for One-dimensional Flow}. {\em \apj\/}
  {\bf 442}, 228.

\bibitem[Schlickeiser {\em et~al.\/}(2011)Schlickeiser, Michno, Ibscher, Lazar
  \& Skoda]{SchlickeiserPRL2011}
{\sc Schlickeiser, R., Michno, M.~J., Ibscher, D., Lazar, M. \& Skoda, T.} 2011
  Modified temperature-anisotropy instability thresholds in the solar wind.
  {\em Phys. Rev. Lett.\/} {\bf 107}, 201102.

\bibitem[{Silva} {\em et~al.\/}(2021){Silva}, {Afeyan} \&
  {Silva}]{SilvaPRE2021}
{\sc {Silva}, T., {Afeyan}, B. \& {Silva}, L.~O.} 2021 {Weibel instability
  beyond bi-Maxwellian anisotropy}. {\em \pre\/} {\bf 104}~(3), 035201.

\bibitem[Thorne \& Blandford(2017)]{TB2017}
{\sc Thorne, K.S. \& Blandford, R.D.} 2017 {\em Modern Classical Physics:
  Optics, Fluids, Plasmas, Elasticity, Relativity, and Statistical Physics\/}.
  Princeton University Press.

\bibitem[Weibel(1959)]{Weibel}
{\sc Weibel, E.~S.} 1959 Spontaneously growing transverse waves in a plasma due
  to an anisotropic velocity distribution. {\em Phys. Rev. Lett.\/} {\bf 2},
  83.

\bibitem[{Wu}(2003)]{Wu2003}
{\sc {Wu}, C.~C.} 2003 {MKDVB and CKB Shock Waves}. {\em \ssr\/} {\bf 107}~(1),
  403--421.

\end{thebibliography}

\end{document}